# Hydrodynamic interaction between coaxially rising bubbles in elasto-visco-plastic materials: Bubbles with a wide range of relative sizes

A. Kordalis[1], Y. Dimakopoulos[1], and J. Tsamopoulos[1,1]

[1] *Department of Chemical Engineering, University of Patras, Patras 26504, Greece*

We consider the buoyancy-driven rise and interaction between two initially stationary, and gravity-aligned bubbles of wide radii ratio and constant volume in an elasto-visco-plastic (EVP) material, extending our previous work regarding bubbles of equal radii [Kordalis *et al*., Phys. Rev. Fluids 8, 083301, (2023)]. Primarily we consider a 0.1% aqueous Carbopol solution and model its rheology with the Saramito-Herschel-Bulkley constitutive model. Initially, we investigate the dynamics for a specific initial separation distance in a wide range of bubble radii and we determine the conditions leading to three distinct patterns: bubble approach, bubble separation and establishment of a constant distance between them. Specifically, when the leading bubble (LB) is smaller than the trailing bubble (TB), the bubbles approach each other due to the smaller buoyancy of the leading bubble. Strong attraction also occurs when the ratio of buoyant over viscous force of both bubbles is considerable. On the other hand, when the size of the TB is such that this ratio is moderate or small, the pattern is dictated by the size of the LB: A significantly larger LB compared to the trailing one causes separation of the pair. On the contrary, an only slightly larger LB may result in the bubbles rising with the same terminal velocity establishing a constant distance between them, the magnitude of which is mainly determined by the elastic response of the surrounding medium. The coupling of a negative wake behind the LB with a slight modification of the stresses exerted at its rear pole generates this dynamic equilibrium. The same equilibrium may be achieved by other specific pairs of bubble sizes for different initial distances of the pair, if a critical initial distance is exceeded. Below this critical value, the bubbles approach each other. Finally, we construct maps of the three patterns with trailing bubble radius versus bubble radii ratio for different initial separation distances and material properties.

---

[1] Corresponding author: tsamo@chemeng.upatras.gr

# I. INTRODUCTION

The dual nature of yield stress materials, i.e. either solid or liquid depending on the imposed stress level, attributes to them a rather complex behavior in most flows. Understanding the way these materials respond affects a very large number of industrial fields and physical processes that these materials are involved in. For example, the food and hygiene industries process mayonnaise, ketchup, chocolate or baby foods [1] and hand gels, creams or toothpaste [2], respectively, while the petroleum and construction industries extract or transport crude oil [3] and produce or handle fresh concrete or plaster [4], respectively. It is evident from the aforementioned examples that the "umbrella" of yield stress fluids contains a variety of very diverse materials that exhibit a macroscopic yield stress. Bubbles are often added to these materials to improve their properties (texture of chocolate) or facilitate their processing and increase their value (cosmetic creams). On the contrary, bubbles must be removed from various materials because they reduce their mechanical properties (fresh concrete) or may cause environmental hazards (nuclear waste slurries, tailings ponds) [5][6]. Determining the conditions for maintaining the bubble distribution and entrapping them in yield stress materials requires studying the dynamics of interacting bubbles [7].

The motion of single bodies driven by buoyancy in non-Newtonian fluids has been studied rather extensively. We will limit our review to the most relevant ideas for the present study. The normal stresses generated in fluids that exhibit elasticity as in viscoelastic or elastoviscoplastic materials, impact the dynamics, and create qualitative changes to the flow with respect to the Newtonian case. For instance, if the moving body is deformable, i.e. a bubble or a drop, then it often acquires the inverted teardrop shape. In the case of bubbles, this shape has been experimentally observed and numerically predicted in viscoelastic fluids [8]–[12], as well as in yield stress fluids [13]–[15]. Furthermore in the case of drops, it has been observed and predicted in viscoelastic fluids [16]–[18] and observed in experiments with yield stress fluids [19]. The common material property giving rise to this shape in deformable bodies is the elasticity of the surrounding medium. Irrespective of the deformability or not of the body, a negative wake can appear behind its rear pole, translating the stagnation point downstream in the middle of the flow. Hence, along the axis of symmetry and closer to the body the velocity direction is the same as the direction of the body translation but becomes opposite to it downstream from the stagnation point. This, in turn, generates two recirculation regions in opposite directions above and below the stagnation point. The overall shape of the negative wake resembles a cone and was originally given this name by Hassager [20], who observed it at the rear of a bubble rising in a viscoelastic fluid. Since then, the negative wake has been observed and predicted behind bubbles in viscoelastic fluids [10][21] and in yield stress fluids [15][22]. The first report of a negative wake, prior to acquiring its name, was in a settling sphere in a viscoelastic solution [23]. However the mechanism producing it was proposed later [24]–[26] and relies on the interplay between shear and extensional stresses. A few years later, it was observed also behind spheres settling in yield stress fluids [27]–[29].

The gravity driven motion and interaction of bodies in Newtonian and, even more so, in non-Newtonian fluids is significantly less understood. The interaction of two equal bubbles ascending in shear-thinning but inelastic liquids was examined experimentally, and complemented by numerical simulations, assuming either axial symmetry for tandem bubbles or 2D flow otherwise [30]. Smaller bubbles initially approached, touched and then rose while remaining close, forming the so-called stable doublet. This behavior is more pronounced in stronger shear-thinning liquids. On the contrary, larger bubbles initially approached, touched and then separated while maintaining a horizontal configuration. The respective behavior in Newtonian fluids is the drafting-kissing-tumbling (DKT) process, during which the pair reaches the horizontal configuration, and the bubbles proceed to separate horizontally. In a shear-thinning liquid the pair is able to reach a steady horizontal arrangement avoiding the 3$^{rd}$ step of the process, depending on the degree of shear thinning it exhibits. The pair interaction is affected by inertia, bubble deformability and initial angle with respect to gravity. When the liquid is both shear-thinning and viscoelastic, tandem bubbles of the same size and below the critical volume for the velocity jump [10] approach each other and merge, while above this volume they can undergo the so-called drafting-kissing-dancing (DKD) process, in which the two bubbles in the 3$^{rd}$ step interchange their positions almost periodically [31]. In elastoviscoplastic materials, tandem bubbles approach each other when they are of equal and sufficient size for buoyancy to overcome the yield stress and set them in motion individually. This occurs irrespective of the initial separation distance, bubble radii or other material properties, because the material is softened by the stresses generated by the leading bubble and retained for a long time facilitating the motion of the trailing one [7]. The velocity jump discontinuity has not been observed or predicted with isolated rising bubbles in EVP materials, as opposed to viscoelastic fluids [32]. Drops in yield stress fluids approach when the initial separation distance is smaller than a threshold, otherwise they behave like isolated drops [19][33][34].

The interaction of solid particles in non-Newtonian liquids has been studied more extensively than that of bubbles and drops. Early experiments [35] demonstrated that spheres attract each other in viscoelastic liquids when their velocities are less than a critical value, whereas they always repel in Newtonian liquids, when the remaining conditions are the same. Moreover, their side-by-side arrangements are unstable because they rotate to become tandem and then precipitate, making the latter arrangement preferred over the former. This study was followed by 2D simulations of falling "disks" in [36], which predicted that when the disks precipitated as a tandem doublet, they approached each other if their original distance was not too large and when set side-by-side they first rotated until the line of centers was aligned with the fall direction and precipitated. More recent studies of two equal spheres coaxially precipitating in viscoelastic and shear thinning fluids determined a critical separation distance below which they approach. Above this critical distance either they behave like isolated particles [37][38] or they separate [38][39], as if this critical distance is a "point of repulsion". The latter case was achieved by changing the solvent viscosity [33] and is rather intriguing. On the contrary, other experimental studies do not identify a critical separation distance and all experiments lead to approach [40][41]. A very nice experimental arrangement was developed by Bot *et al*. [42] to initiate the motion of two tandem spheres simultaneously, instead of sequentially, which is typically done. They reported the existence of a critical separation distance between two equal spheres in a Boger fluid: spheres closer to each other or further away than this distance moved to attain it. Hence this distance constitutes a steady separation distance or a "point of attraction" for the two spheres. The flow initiation in [42] allows the simultaneous acceleration of the spheres, which may reach a steady velocity in a "virgin" material, whereas the more often used sequential initiation allows the leading sphere to accelerate before the trailing one is even allowed to move through the material the properties of which may have been modified by the passage of the leading sphere. Perhaps because of the very diverse and not fully reported or controlled conditions in this larger number of studies on interacting spheres, the results are not always fully justified and are often conflicting. A comprehensive review of these and other phenomena arising when bubbles, drops or particles interact under gravity but in the absence of external flow in viscoelastic and shear-thinning liquids appeared recently [43].

Clearly, the motion, interaction and deformation of particles may be affected by viscous, inertial, elastic, plastic, and capillary forces, and the material functions that affect them. It also depends on the particle deformability, their relative position with respect to the gravitational field, the distance between them, which is related to the concentration of the dispersed phase, and the size distribution. This large number of parameters results in dimensionless numbers that are too many to examine them in full, necessitating each study to focus on a well-defined window of relevance to the application in mind and to precisely report their values. The original assumptions that the particles are and remain spherical, and are of equal diameter, that the flow is creeping, or potential were required to derive analytical solutions or use older numerical methods of solution. These and other similar restrictions can now be avoided with the development of accurate interface capturing methods such as the volume of fluid method (VOF), a good example of which is Basilisk [44] or interface tracking methods, such as the arbitrary Lagrangian- Eulerian- (ALE), a good example of which is PEGAFEM-V [45][46]. The former can more easily simulate particle deformation and merging even in three dimensions, while the latter can more accurately simulate flows in non-Newtonian fluids. Recently, the former has been used by Zhang et al. [47][48] to simulate the dynamics of a pair of equal and deformable bubbles under moderate or high inertia in Newtonian liquids and revealed a wealth of possible interaction scenarios. The latter has been used to examine the rise and interaction of a single or a pair of bubbles in an elastoviscoplastic liquid [7][15] and the sedimentation of a charged solid particle in a viscoelastic liquid [49], but also in several other moving boundary or confined flows.

In the present study we examine the in-line configuration of bubbles allowed to rise simultaneously in an elastoviscoplastic material due to buoyancy. Choosing this tandem arrangement is justified by the earlier findings [7] that the translation of the leading bubble softens the material, which remains so for very long time creating a corridor for the trailing bubble to move into. This is similar to the ideas of the corridor of reduced viscosity set forth in [35]. The material under study has been characterized via shear experiments to determine its rheological properties. We use PEGAFEM-V [45][46] to simulate various combinations of different radii for the leading and the trailing bubble, maintaining their volume constant during the rise. We identify conditions leading to bubble approach, separation or reaching an equilibrium distance. We explain the forces leading to this equilibrium when the bubbles are unequal and evaluate the slightly repulsive role of the negative wake. We assemble maps recording the behavior of each bubble pair for a certain value of the initial separation distance, and then for additional values of the initial separation distance and the more important material properties.

## II. PROBLEM FORMULATION

We consider the buoyancy-driven rise of two bubbles with a wide range of relative sizes, positioned in line with the gravity field in an unbounded elasto-visco-plastic (EVP) material. The bubbles maintain their constant volume throughout the simulation. The quantities bearing the superscript * are dimensional, otherwise they are dimensionless. At time $t^* = 0$ the bubbles have an initial separation distance $d_o^*$ between their geometric centers. The material is incompressible with constant density $\rho^*$, elastic modulus $G^*$, consistency index $k^*$, shear-thinning exponent $n$ and yield stress $\tau_y^*$. We employ the Saramito-Hershel-Bulkley (SHB) [50] constitutive model to capture the rheology of the material. The density and viscosity of the gas bubble are assumed to be negligible in comparison to the corresponding properties of the material. We also consider that the surface tension $\sigma^*$ of the fluid-gas interface is constant in space and time. The bubbles are simultaneously set free from rest under the effect of buoyancy. We adopt a cylindrical coordinate system with $\{r^*, z^*, \theta\}$ the radial, axial, and azimuthal coordinates, respectively, and assume axial symmetry. The center of the coordinate system is placed in the middle of the instantaneous distance between the two bubble centers and moves upwards with a distinct velocity $U_m^*$. The gravity vector points to the negative z-direction. The bubble above the coordinate center will be referred to as the leading bubble (LB) and the bubble positioned below it as the trailing bubble (TB) with the radii of their initial spherical shapes being $R_{LB}^*$ and $R_{TB}^*$, respectively.

We scale all lengths with the radius of the TB, $R_{TB}^*$, the velocities by balancing buoyancy with inertial forces, $\sqrt{g^* R_{TB}^*}$, and time with a convective scale, $\sqrt{R_{TB}^*/g^*}$. Consequently, we arrive at the dimensionless numbers that govern this problem:

$$Ar = \frac{\rho^* g^* R_{TB}^*}{k^* \left(\frac{\sqrt{g^* R_{TB}^*}}{R_{TB}^*}\right)^n}, \quad Bn = \frac{\tau_y^*}{\rho^* g^* R_{TB}^*}, \quad Bo = \frac{\rho^* g^* R_{TB}^*}{\frac{\sigma^*}{R_{TB}^*}}, \quad Eg = \frac{\rho^* g^* R_{TB}^*}{G^*}, \quad \delta = \frac{R_{LB}^*}{R_{TB}^*} \quad (1)$$

$Ar$ is the Archimedes number that balances buoyancy to viscous force, $Bn$ is the Bingham number that balances plastic to buoyancy force, $Bo$ is the Bond number that scales buoyancy with the capillary force. The last two are $Eg$, the elastogravity number, which scales buoyancy forces to elastic forces and the size ratio $\delta$ that compares the radius of the LB with the radius of the TB.

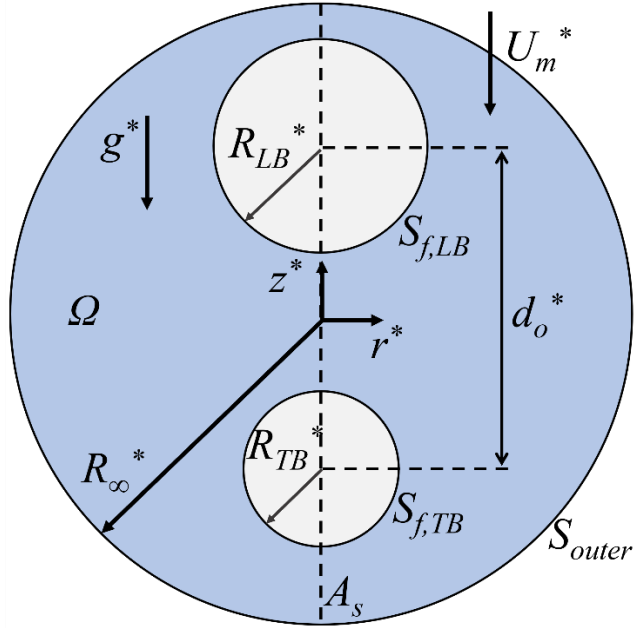

FIG.1 Schematic representation of the two coaxial bubbles of a given size ratio in an EVP material at the initial rest state. Here $\Omega$ denotes the volume of the EVP material, $S_{f,TB}$ and $S_{f,LB}$ the interfaces between air and the material for the trailing bubble and leading bubble, respectively, $A_s$ and $S_{outer}$ the symmetry axis and the outer boundary respectively. The coordinate system is placed at the middle of the instantaneous distance between the bubble centers.

Using the aforementioned nondimensionalization we arrive at the dimensionless forms of momentum and mass conservation equations:

$$\left(\frac{\partial \boldsymbol{u}}{\partial t} - \frac{dU_m}{dt}\boldsymbol{e}_z\right) + \boldsymbol{u} \cdot \nabla \boldsymbol{u} = \nabla \cdot \boldsymbol{T} - \boldsymbol{e}_z \quad (2)$$

$$\nabla \cdot \boldsymbol{u} = 0 \quad (3)$$

Where $T$ is the Cauchy stress tensor split into pressure and extra stress $T = -PI + \tau$. The fact that the coordinate system moves with the bubbles makes it non-inertial and requires the extra term $\frac{dU_m}{dt}e_z$ [24]. The datum pressure is set to zero at z=0 on $S_{outer}$. When motion is initiated, the coordinate system translates upwards, following the bubbles. We keep a constant volume throughout the simulations for each bubble as explained in [7]. The extra stress tensor is related to the rate of strain, $\dot{\gamma}$, according to the SHB model:

$$Eg\overset{\triangledown}{\tau} + max[Ar(|\tau_d| - Bn), 0]^{\frac{1}{n}}\frac{\tau}{|\tau_d|} = \dot{\gamma}, \tag{4}$$

where $\tau_d = \tau - tr(\tau)/tr(I)$ is the deviatoric part of the stress tensor and $|\tau_d| = \sqrt{0.5\tau_d:\tau_d}$ is its magnitude. The max term in Eq. (4) introduces the von Mises criterion and dictates whether the material is yielded or not. When $|\tau_d|$ is smaller than $Bn$ the max term vanishes, and hyperelastic solid behavior is followed. Otherwise, the material flows. The yield surface, the interface between unyielded and yielded material, arises at $|\tau_d| = Bn$ and is determined *a-posteriori*. The upper-convected time derivative is defined as:

$$\overset{\triangledown}{\tau} = \frac{\partial \tau}{\partial t} + u \cdot \nabla \tau - (\nabla u)^T \cdot \tau - \tau \cdot \nabla u \tag{5}$$

On the surface of each bubble a local force balance holds between the total stress tensor of the fluid, the pressure of each bubble and capillarity:

$$-n_i \cdot T = P_{b,i}n_i + Bo^{-1}(2\mathcal{H}_i)n_i, \quad \text{on } S_{f,i} \text{ where } i = LB, TB \tag{6}$$

Where $n_i$ denotes the material-outward, unit, normal-vector to the free surface of each bubble, $2\mathcal{H}_i = \nabla_s \cdot n_i$ is twice the mean curvature of a free surface with $\nabla_s = (I - nn) \cdot \nabla$ the surface gradient operator. $P_{b,i}$ is the pressure in bubble $i$. Also, we impose the kinematic boundary condition on each bubble surface:

$$n_i \cdot \left(u - \frac{\partial r_f}{\partial t}\right) = 0, \quad \text{on } S_{f,i} \text{ where } i = LB, TB \tag{7}$$

where the derivative in (7) denotes the velocity of the mesh nodes in the fluid domain and $r_f$ denotes the position vector of points on the free surface.

The remaining boundary and initial conditions are: (a) symmetry of velocity and stress components along the geometric symmetry line, (b) the velocity of the outer boundary, located at $r = \sqrt{R_\infty^2 - z^2}$, with $R_\infty = 500$, is $-U_m e_z$, making the center of the coordinate system stationary, and (c) initially the two bubbles are spherical and the fluid is at rest and stress-free. The pressure inside each bubble, $P_{b,i}$, and the rise velocity of the coordinate system $U_m$ are determined by imposing the constant volume of each bubble and that the coordinate system is always positioned at the middle between the centers of volume of the bubbles, as detailed in [7].

### III. NUMERICAL IMPLEMENTATION

These coupled equations are solved numerically using the Petrov-Galerkin stabilized Finite Element Method for Viscoelastic flows (PEGAFEM-V), [45][46]. The novelty of this method is the incorporation of equal order interpolants for all variables by extending the PSPG methodology in [51], in this way improving numerical stability and reducing the computational cost, the effort for code development and the duration of simulations. DEVSS [52] is also used to maintain the elliptic nature of the momentum equation even when the Newtonian solvent is absent. The hyperbolic character of the constitutive model is handled via the SUPG formulation [53]. The weak form of the equations, their implementation in the moving boundary problems and other details can be found in [46]. The validity of the method has been tested thoroughly through several benchmarks, in which it exceeded previous High-Weissenberg number limitations, and solved long-standing problems in viscoelastic flows, such as the sharkskin instability [54], and in more complex flows regarding elastoviscoplastic materials [15][55][56]. Moreover, we employ the arbitrary Lagrangian-Eulerian (ALE) framework to track the liquid-air interfaces and choose the elliptic grid scheme [57], [58] to track the location of the mesh nodes in the fluid domain. When the deformation of the mesh is large and the mesh quality drops below a specified limit, we employ a remeshing algorithm [7] to prevent the accumulation of numerical error. Time integration is carried out using a fully implicit, second order, backward finite difference scheme, accelerated by using as initial guess for the Newton algorithm the quadratic extrapolation of the previous solution. The timestep is set to be constant throughout each simulation and equal to 0.02.

## IV. RESULTS AND DISCUSSION

For most of the present simulations, we have chosen the 0.1% w/w Carbopol solution characterized and employed in bubble rise experiments in [13]. Its preparation followed the usual protocol of adding the polymer in deionized water under continuous mixing and, finally, PH neutralization by NaOH. We fitted the Saramito-Herschel-Bulkley (SHB) model to extract its parameters, knowing that the ratio of the solvent viscosity to the total viscosity of the Carbopol gel to be negligibly small and setting it to zero. The elastic modulus $G^*$ is acquired from the SAOS experiments they provide, while the other three properties, namely the yield stress, $\tau_y^*$, consistency index, $k^*$ and the HB exponent, $n$ are derived by fitting the model to shear rheological data from their experimental study. Details of the fitting process can be found in [7]. The resulting values are given in Table I.

TABLE I. Values of the properties for the 0.1% Carbopol solution of Lopez et al. [13].

| $G^*$ $(Pa)$ | $\tau_y^*$ $(Pa)$ | $k^*$ $(Pa\ s^n)$ | $n$ |
|---|---|---|---|
| 40.42 | 4.8 | 1.75 | 0.47 |

We proceed with the analysis of the transient rise of the two coaxially placed bubbles in the EVP fluid. Since we have already studied extensively the effect of the material properties on the interaction of two equal bubbles in our previous work, we keep the properties of the material fixed to the values given in Table I for most of this study and, in the end, we examine the effect of varying the two more important parameters, $G^*$ and $\tau_y^*$. We divide this section into four parts. In the first part we combine bubbles of various sizes into pairs, and we examine their rise for the base value of the initial separation distance equal to 5. Three possibilities arise: bubble attraction, bubble separation and establishment of an equilibrium distance. Then, we construct a map of these possibilities for this distance. In the second part, we construct the maps for two additional initial separation distances. In the third part, we focus on the conditions leading to the equilibrium separation distance. Specifically, we examine its origin with emphasis on the factors that make it a stable configuration and the effect of the initial separation distance to it. In the fourth part, we assess the slowing effect of the negative wake ahead of the trailing bubble. In the fifth part, we examine the effect of varying $G^*$ and $\tau_y^*$ on the three possibilities. In the simulations we vary the radius of the TB and, hence, the dimensionless numbers that depend on it, while the radius of the LB is introduced independently by varying the dimensionless ratio of radii, $\delta$. Hence, the maps are constructed in terms of the dimensional $R_{TB}^*$ and the dimensionless $\delta$.

### A. Initial separation distance $d_o = 5$

We keep a range of sizes between 3.6 mm and 16 mm for the bubbles, based on the experimental results for single bubbles in the 0.1% Carbopol [13]. Moreover, the plastic behavior of the material dominates below the lower limit of 3.6 mm, according to Kordalis *et al.* [7]. Above the upper limit of 16 mm, the bubbles are large enough to introduce considerable inertia and make the axisymmetric assumption questionable. Single bubbles in the selected size range are known to follow axisymmetric shapes in EVP materials, according to the comparison between experiments and simulations shown in the work of Moschopoulos *et al.* [15].

#### *1. Trailing bubble of 4mm*

In this subsection we explore the dynamics of multiple pairs of bubbles having a TB of radius 4 mm. The size of the LB is such that (A) $\delta = 0.9$, (B) $\delta = 1.04$, (C) $\delta = 1.05$ and (D) $\delta = 1.10$. The dimensionless numbers of cases (A)-(D) depend solely on the $R_{TB}^*$, so in all cases they have the same values, given in Table II.

Table II. Parameter values for cases (A)-(D), with $R_{TB}^* = 4\ mm$

| $R_{TB}^*$ | Ar | Bo | Eg | Bn |
|---|---|---|---|---|
| 4 mm | 3.61 | 2.15 | 0.97 | 0.12 |

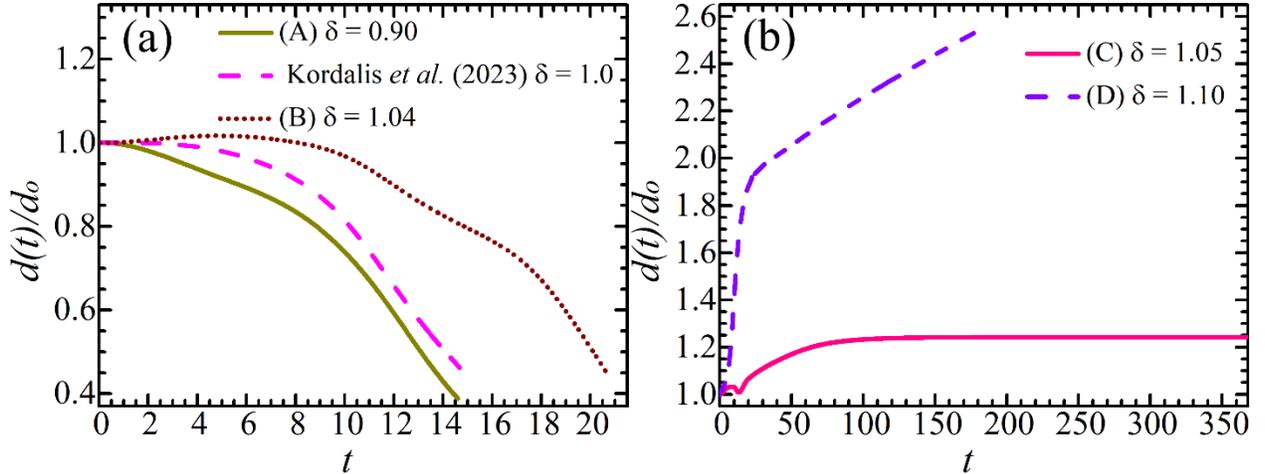

FIG 2 Normalized separation distance, $d(t)/d_o$, versus time in panel (a) for (A) $\delta = 0.90$, the case from Kordalis *et al.* [7] for equal bubbles of 4 mm, and (B) $\delta = 1.04$ showing bubble *approach* of all three pairs. In panel (b) we depict $d(t)/d_o$ for (C) $\delta = 1.05$ showing the formation of the *equilibrium distance* and (D) $\delta = 1.10$ showing the *separation* of the pair. The steady separation distance, $d_s$, is generated only in case (C).

In Fig. 2 (a) we depict the time evolution of the normalized separation distance of cases (A), and (B) along with the equal bubble configuration from Kordalis et al. [7], all predicting approach. In Fig. 2 (b) we show cases (C) and (D) predicting either establishment of a steady distance or separation of the pair respectively. Starting with Fig. 2 (a), we observe that the pair (A) approaches faster than the equal bubble pair. This is expected since a smaller LB generates a smaller driving force (buoyancy), and the larger TB catches up with it faster. On the other hand, we observe that the pair (B) manages to approach even though its LB is larger than its TB, although the approach is slower compared to the equal size pair. The approach occurs because the somewhat larger buoyancy of the LB is counterbalanced by the decreased drag exerted on the TB. This decrease is caused by two factors: (a) the previous passage of the LB creates stresses, which remain stored in the material, making it softer and (b) the shear and extension thinning of the material also contribute to the creation of corridor of reduced viscosities. For an extended analysis on the mechanism of approach and the transient drop of the drag on the TB the interested reader is referred to Kordalis et al. [7]. Increasing the size ratio by only 1% to $\delta = 1.05$ (case (C)) we observe that after an initial transient period the distance between the bubbles asymptotes to a constant value. This relatively small increase in the buoyancy of the LB balances exactly the decrease of drag on the TB and the tandem bubble configuration reaches a constant rising velocity, which is the same for both bubbles, hence the steady separation distance appears. Henceforth, we will refer to the constant distance formed as steady separation distance, or steady distance or equilibrium distance and denote it as $d_s$. We will discuss more thoroughly the formation of the steady separation distance in section IV.C. Also, in Fig. 2 (b) we see the time evolution of the normalized separation distance for case (D), which indicates separation of the pair. After an initial transient period of abrupt increase in the distance between the bubbles, the slope changes to a slower linear increase with time. Both bubbles have acquired their final shapes, and they rise with distinct steady velocities with the LB having a higher velocity than the TB.

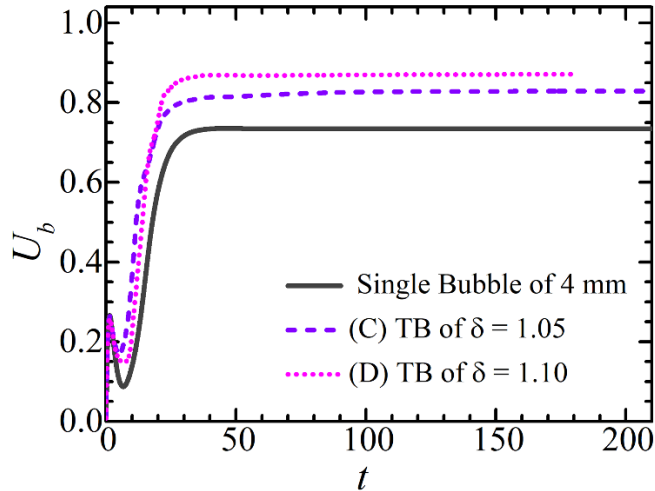

FIG 3 Time evolution of the bubble velocity for the TB with $R_{TB}^* = 4\ mm$ tandem with the larger leading bubble of cases (C) and (D) compared with a single isolated bubble of the same radius.

In Fig. 3 we depict the time evolution of the velocity of the TB from cases (C) and (D) juxtaposed with the velocity of a single rising bubble with the same radius. The larger the LB is, the larger the steady velocity acquired by the TB of the same radius, $R_{TB}^* = 4\ mm$. The TB of case (C) shows a velocity increase of 13% with respect to the single bubble while the TB of case (D) shows an increase of 18%. The same mechanisms responsible for the reduced drag exerted on the TB induce a larger velocity to the TB. We can elaborate on that through the definition of the drag coefficient if we assume that the medium exhibits an equivalent constant viscosity $\eta^*$ of a Newtonian fluid. For this rough estimation we use the Stokes drag coefficient of a spherical bubble, given that the Archimedes number $Ar$ is $O(1)$.

$$c_d = \frac{16}{Ar} = \frac{8}{3U_b^2}, \qquad (14)$$

where $U_b$ is shown in Fig. 3 and is the steady dimensionless velocity of the TB we examine. The second equality of Eq. (14) stems from the definition of drag coefficient when steady conditions are reached [15]. $Ar$ can be reformulated to create a characteristic viscosity that we compare among the cases.

$$Ar = \frac{\rho^* \sqrt{(g^* R_{TB}^*)}^2}{\eta^* \frac{\sqrt{(g^* R_{TB}^*)}}{R_{TB}^*}} = \frac{\rho^* \sqrt{(g^* R_{TB}^*)} R_{TB}^*}{\eta^*} \qquad (15)$$

Combining Eq. (14) and (15) we obtain an expression for the apparent viscosity that the TB experiences when the steady conditions are reached:

$$\eta^* = \frac{\rho^* \sqrt{(g^* R_{TB}^*)} R_{TB}^*}{6\, U_b^2} \qquad (16)$$

According to this rough estimation, the TB of case (C) experiences an apparent viscosity of the surrounding fluid $\eta^* = 0.192$ Pa s, for the TB of case (D) $\eta^* = 0.174$ Pa s, while the single bubble feels an equivalent viscosity of $\eta^* = 0.245$ Pa s.

In Fig. 4 we show snapshots of the bubble pair shapes accompanied by contours of the shear and axial normal stresses in the fluid for cases (A)-(D). We also include the yield surface in each plot. We start from case (A) in Fig. 4 (a) at the initiation of the flow. The yielded area consists of three distinct parts, the first being the area of high stresses surrounding the front of the LB. We observe that the size of this yielded area is smaller compared to the corresponding yielded area for larger LBs, in Fig. 4 (e), (i) and (m), all at $t = 3$. This denotes that the smaller LB is less effective in breaking the material ahead of it. The second yielded part is between the two bubbles and follows the spatial distribution of the shear stress contours. In the middle there is an island of unyielded material since around this area the extensional stresses go through zero as they vary from the extensional condition behind the LB to the compressive one in front of the TB and the shear stress has small values close to the axis of symmetry. Finally, there is the yielded area behind the TB. The same yielded areas appear in the early snapshots of all cases ($t = 3$). Similar reasoning exists for the yielded areas regarding the rest of the snapshots. Continuing to Fig. 4 (b), the elastic normal stresses elongate the TB but not the LB, which assumes a rather compressed shape. In Fig. 4 (c) the separation between the bubbles has decreased and the shape of the LB is compressed even more. The compression of the LB is maximized in Fig. 4 (d) when the bubbles are nearly touching. The compressed shape of the LB is a result of the presence of the TB and cannot be observed in isolated rising bubbles of this size in this medium. The smaller driving force and velocity of the LB results in the shrinking of the free space downstream of it, which is necessary for tensile stresses to deform the LB to a more prolonged shape.

For case (B), the bubbles have started to move in Fig. 4 (e) and in Fig. 4 (f) they have started to deform. Now the LB becomes prolate in contrast to what we observe in Fig. 4 (b). In Fig 4 (g) it acquires an inverted teardrop shape with a rounded rear. The formation of this hydrodynamic shape prologs the time for approach to occur. Finally, the pair comes close in Fig. 4 (h). The effect of the transient development of the tensile stresses is visible by comparing Fig. 4 (d) and Fig. 4 (h). In the first case the shape of the LB deforms only slightly in the z-axis and due to the pushing from the TB it resembles a shape that exhibits mixed elastic-inertial features, on which we will elaborate in the next section. In the second case, the shape of the LB is prolate with a flat rear. At all times and for both cases, the bridge of shear stresses connecting the back of the LB to the front to the TB is present.

In case (C), the bubbles have started rising in Fig. 4 (i) and are deformed in Fig. 4 (j). Then in Fig. 4 (k), they increase their separation distance by a small amount, and both assume the inverted teardrop shape, since the tensile stresses have developed at both their rear poles. Then, at a very long time ($t = 350$) they reach the steady distance configuration shown in Fig. 4 (l). The shear bridge is still present, and the bubbles are englobed in one big, yielded envelope. In Fig. 4 (k) and (l) as well as (h) we observe a sharp yielded edge behind the TB. The shear stresses there are small, but the extensional ones remain and are mainly responsible for the yielding of the material there. These areas are affected by the presence of the negative wake in this region.

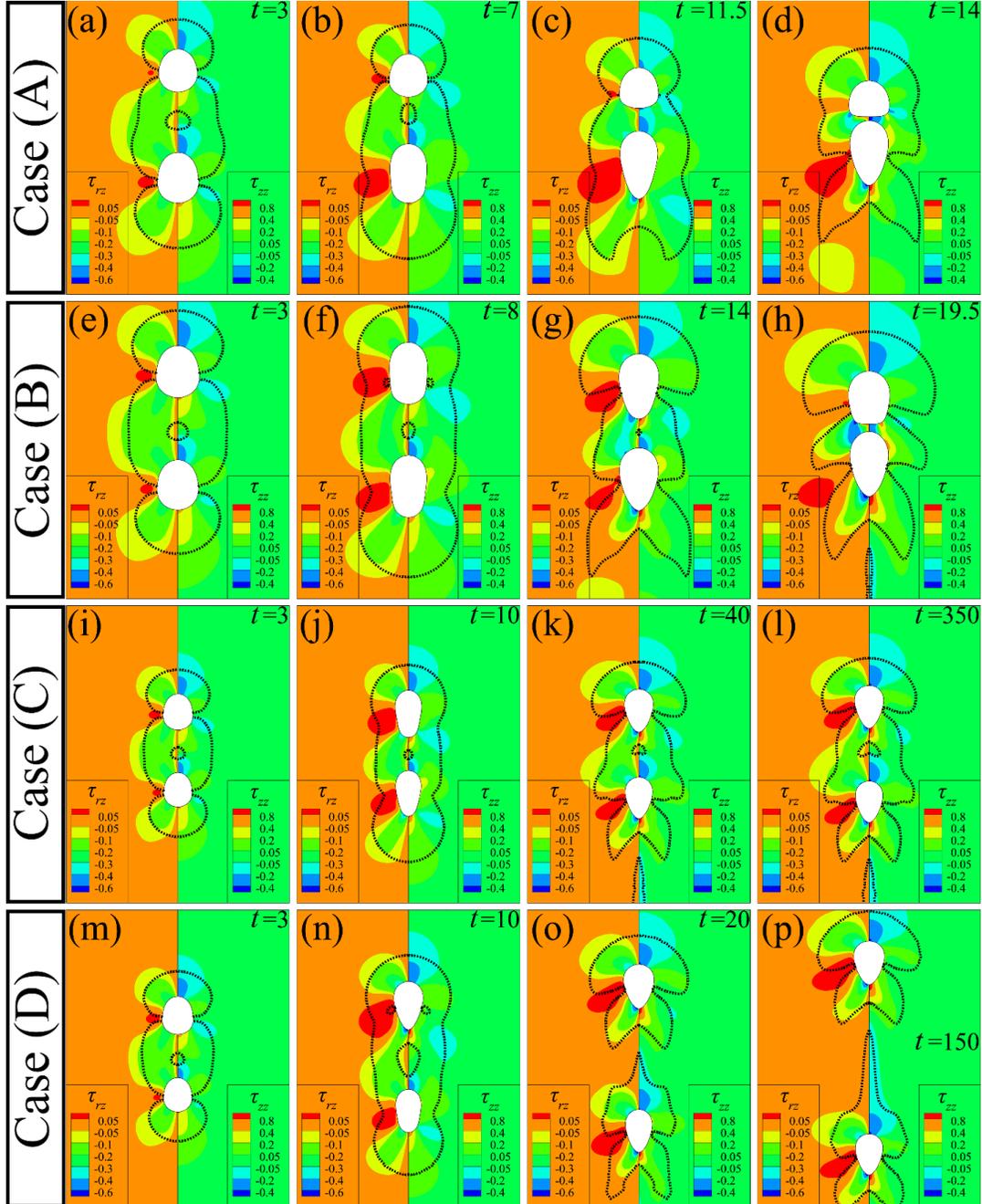

FIG 4 Snapshots during the rising of each bubble pair at the indicated times. Each panel shows the shear stresses in the left half, the normal axial stresses in the right half along with the yield surface depicted with a dotted black line. Panels (a)-(d) correspond to case (A), panels (e)-(h) correspond to case (B), panels (i)-(l) correspond to case (C) and panels (m)-(p) correspond to case (D). Note that snapshots of cases (C) and (D) are zoomed out, hence the bubbles seem smaller.

In case (D), the bubbles have started moving coaxially in Fig. 4 (m) and have deformed already in Fig. 4 (n). The LB has obviously assumed the inverted teardrop shape earlier than in (C) due to its larger buoyancy and starts increasing the distance separating it from the TB already from $t = 4$ (n). That explains why the separation distance increases abruptly at early times in Fig. 2 (b). The TB needs more time to create a hydrodynamically favorable shape than the LB and so, in the meantime, the LB has drifted apart. In Fig. 4 (o) the separation intensifies as the shear bridge connecting the bubbles breaks. Consequently, the yielded area between the bubbles also breaks into two parts, one following the rear pole of the LB and the other one staying with the front pole of the TB. In this time instant we see the formation of a yielded tip at the front yielded area of the TB. The yielded tip grows and in Fig. 4 (p) it is quite large since the separation distance is the largest from all snapshots shown. Its presence is caused by extensional stresses, so it has the same origin as the sharp yielded edge we discussed earlier for panels (k) and (l), except that it is formed behind the LB. The areas are affected by the presence of a negative wake there.

### *2. Map of regimes*

Next, we assemble a map of regimes distinguished by the three outcomes of bubble interaction: *approach*, *separation*, and establishment of an *equilibrium distance* in the parametric space of $R^*_{TB}$ vs. $\delta$ for a single initial distance, $d_o = 5$. We showed in the previous subsection, IV.A.1, the existence of these three regimes, and now we gather our results for multiple combinations of the LB and the TB sizes.

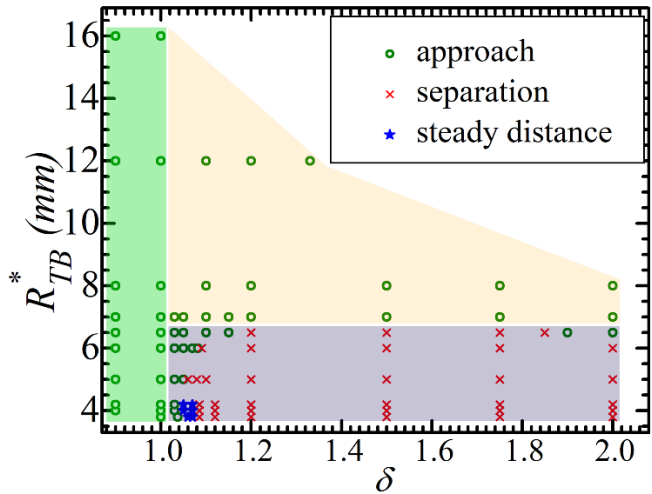

FIG 5 Map of regimes for multiple combinations of $R^*_{TB}$ and $\delta$ at $d_o = 5$. We depict approaching bubbles with green circles, formation of equilibrium distance with blue stars and separating bubbles with red crosses. The area with green color (first area) corresponds to the approach induced by the buoyancy difference between the bubbles, the area with light yellow color (second area) corresponds to the approach due to large inertia. The third area indicated by the purple color includes all three patterns.

In Fig. 5, each point with coordinates $(\delta, R_{TB}^*)$ corresponds to a unique complete transient simulation and is indicated by a mark of different color and shape. The three different marks indicate the three final outcomes: approach, separation or establishment of a constant distance. However, to facilitate our discussion we introduce three areas indicated by different colors and characterized by different dominant forces. The first area refers to a LB smaller or equal compared to the TB, $\delta \leq 1$. These cases are enclosed by the two vertical lines of $\delta = 0.9$ and $\delta = 1$ in Fig. 5, forming the domain in green color, where only approaching occurs. Due to its smaller total force, i.e. the buoyancy minus the drag, the LB cannot rise faster than the TB. This difference is reinforced by a combination of decreased drag on the TB and the lack of a hydrodynamically favorable shape of the LB because of the limited space between the bubbles [7]. Because of these reasons leading to bubble approach, approach will also take place even for smaller $\delta$, not shown in Fig. 5.

In the second and third distinct areas, the LB is larger than the TB ($\delta > 1$). The second area is occupied by pairs with $R_{TB}^* \geq 7\ mm$, which exhibit only approach. At this point, we should recall an observation from our previous study on equal bubbles [7]. The response of bubbles with radius $R_{TB}^* \leq 6\ mm$, is elastoplastic, meaning that elasticity originating from the solid response of the material builds the tensile stresses, which in turn deforms the bubbles to more prolate shapes. The dimensionless number that quantifies the impact of elasticity is the auxiliary Weissenberg number, defined as:

$$Wi = \frac{k^*}{G^*}\left(\frac{g^*}{R_{TB}^*}\right)^{n/2} = \frac{Eg}{Ar} \tag{17}$$

On the contrary, the response of the system when bubbles have $R_{TB}^* \geq 8\ mm$ is mainly inertial. This means that the larger buoyancy of the bubbles leads to significant inertia of the material and oblate bubble shapes are formed. The dimensionless number that quantifies the impact of inertia is $Ar$. Consequently, for $6\ mm \leq R_{TB}^* \leq 8\ mm$, the response of the system is mixed, elastoplastic and inertial. At $R_{TB}^* = 6\ mm$, elastoplasticity is primary and inertia secondary. On the contrary, at $R_{TB}^* = 8\ mm$ the opposite is true, and inertia is the primary factor among the two. Keeping this in mind, we observe in Fig. 5 that only approaching is observed for $R_{TB}^* \geq 7\ mm$. This area is colored light yellow and extends to even larger $R_{TB}^*$ than indicated in the figure as long the phenomenon remains axisymmetric. Even if the radius of the LB is twice as large as the TB radius ($\delta = 2$), the pair comes close eventually. This result is quite unexpected, if we consider that the driving force of motion for the LB is $\delta^3$ times larger than the respective one for the TB. However, when inertial forces cannot be considered small, the generated vorticity affects the pressure field around the surface of a bubble, dropping the dynamic pressure at its rear pole [59]. The term dynamic pressure refers to the pressure developed due to flow without the hydrostatic part, defined in dimensionless form in Eq. (18).

$$\wp(r,z) = P(r,z) + z \tag{18}$$

So, when tandem bubbles are examined, the TB experiences a reduced dynamic pressure at its front due to the vorticity generated by the LB. The result is an attraction of the TB towards the LB, a phenomenon called sheltering effect or drafting [47]. Drafting is further promoted when the bubble shapes are more oblate, spatially shielding more the TB from the flow. This denotes a dependence on $Bo$, the value of which dictates how easy it is to deform the shape of a bubble. According to the extensive work of Zhang *et al.* [47] on the interaction of tandem equal-sized bubbles in a Newtonian fluid, easier deformability of the bubbles, or equivalently a larger $Bo$, aids to the approach of the pair.

To understand and reveal the mechanism resulting in approach for $R^*_{TB} \geq 7\ mm$ we will examine cases (E) and (F). What is special about these two cases is that the behavior of the system changes drastically with a minor increase in the size of the LB, with (E) $\delta = 1.85$ and (F) $\delta = 1.90$. Indeed, along the horizontal line of $R^*_{TB} = 6.5\ mm$ in Fig. 5, we observe that by this small increase in $\delta$, the response transitions from separation to approach. This means that in front of the same TB, a smaller LB results in separation of the pair, while a larger one results in approach of the pair. Since the radius of the trailing bubble is the same, $R^*_{TB} = 6.5\ mm$, both cases have the same parameter values, which are given in Table III:

Table III. Parameter values for cases (E)-(F), with $R^*_{TB} = 6.5\ mm$

| $R^*_{TB}$ | $Ar$ | $Bo$ | $Eg$ | $Bn$ |
|---|---|---|---|---|
| 6.5 mm | 6.57 | 5.68 | 1.58 | 0.075 |

In Fig. 6 we show snapshots of cases (E) and (F) at the same time instants. As we can see from Fig. 6 (a), the LB of case (F) is only slightly larger than the one of case (E). At the initial time instants Fig. 6 (a) and Fig. 6 (b), the contours of the dynamic pressure are almost identical between the two cases. During these early times, the flow field evolves, but no significant distinction can be made either in the shapes of the respective bubbles or in the spatial distribution of the dynamic pressure. However, in Fig. 6 (c), the bubble pair in (E) starts to separate while the bubble pair in (F) maintains the same distance as before. The dynamic pressure at the front of the TB in (E) increases significantly, raising the resistance to its movement, while in (F) it remains the same with the earlier instant. In (F), the slightly smaller distance of bubbles and the higher inertia due to the larger size of the LB bends it more, sheltering more effectively the TB against the downward flow and hence it allows the TB to be drawn towards the LB even more, accelerating the process. In Fig. 6 (d), the pair of (E) has increased its distance by a factor of more than two while the bubbles of (F) have approached each other and are about to touch.

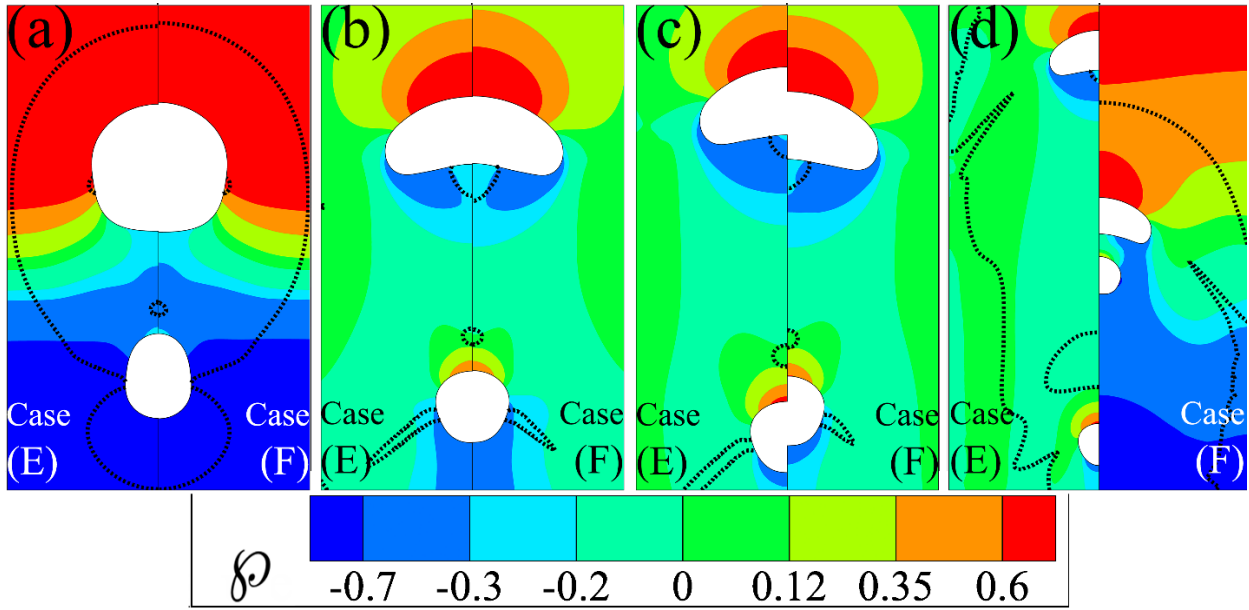

FIG 6 Comparison of the dynamic pressure developed for a 6.5 mm TB of cases (E) $\delta = 1.85$ to the left and (F) $\delta = 1.90$ to the right of each panel at time (a) 3, (b) 14, (c) 34, and (d) 88. We also superimpose the yield surface with black dashed line. Note that snapshots in (d) are zoomed out, hence they seem smaller.

Drafting dominates the interaction when inertia is important enough and favors the approach of the bubbles. From Eq. (17) $Ar$ is inversely proportional to $Wi$. According to our predictions, when $Ar > 6.57$ (or $Wi < 0.24$), drafting prevails at the present initial separation distance and only *approach* is observed. This constitutes the second area of Fig. 5. The elastic response and the inertial response of the fluid are antagonistic, as shown by Eq. (17). The stabilizing elastic response of the fluid is suppressed by the inertia of the fluid, in the sense that the dynamics changes and normal stresses developed due to the flow field cannot deform the bubbles towards prolate shapes. Hence, the shapes of the bubbles become oblate. On the other hand, when $Ar \leq 6.57$ (or $Wi \geq 0.24$), all three patterns appear. This constitutes the third area of Fig. 5. We should underline that although cases (E) and (F) were used to explain the dynamics of approach of the second area where inertia of the fluid is prevalent, the horizontal line of $R_{TB}^* = 6.5\ mm$ is included in the third area of interest since all kinds of regimes are predicted there. However, we consider it as the limit between the two areas. As we will discuss subsequently in our analysis, this threshold value strongly depends on the initial separation distance.

Finally, the third area of Fig. 5 is depicted in purple and it does not indicate a single specific pattern. Along all horizontal lines of the same TB radius we see a transition from *approach*, at smaller $\delta$, to *separation*, at larger $\delta$. An important remark is that at smaller TB radii, in the vicinity of 4 mm, the transition goes through the third possibility, the formation of *steady separation distance*. On the contrary, the transition at larger radii occurs directly. Our simulations do not predict the steady distance formation for example at $R_{TB}^* = 5\ mm$ or $6\ mm$ at the initial separation examined in this section, even though we performed simulations with very fine increases in $\delta$. For instance, in the horizontal line of $R_{TB}^* = 6\ mm$, we varied the value of $\delta$ to the third decimal digit and, yet *no steady distance* was observed.

### B. Maps of regimes for different initial distances, $d_o$

In this section we present simulation results for similar combinations of $(\delta, R_{TB}^*)$ as in section IV.A.2 for two additional initial distances, $d_o$, one below, $d_o = 4$, and the other one above, $d_o = 6$, the base value of $d_o = 5$. The objective is to improve our understanding of the bubble interaction. Due to the large number of simulations that are required to produce each map, we have selected only key values of the TB radius.

In Fig. 7 we show the maps for the three initial separation distances. A common feature of all three panels is that when the LB is smaller than or equal to that of the TB, $\delta \leq 1$, or the ratio of buoyant to viscous forces is large enough, $R_{TB}^* \geq 8\ mm$ (at least up to the sizes we report in these panels), the bubbles always approach each other. The reasons and mechanisms leading to *approach* in both areas of parameter space were presented in the previous section. Moreover, the approach for $R_{TB}^* \geq 8\ mm$ happens as long as the initial separation distance is small enough compared to the size of the wake generated by bubbles in this range of radii. Generally in Fig. 7 (a), bubble pairs approach each other for the most part not attaining a steady final distance, whereas in Fig. 7 (c) an area occupied by steady distance pairs exists and is larger than that in the base case (Fig. 7 (b)).

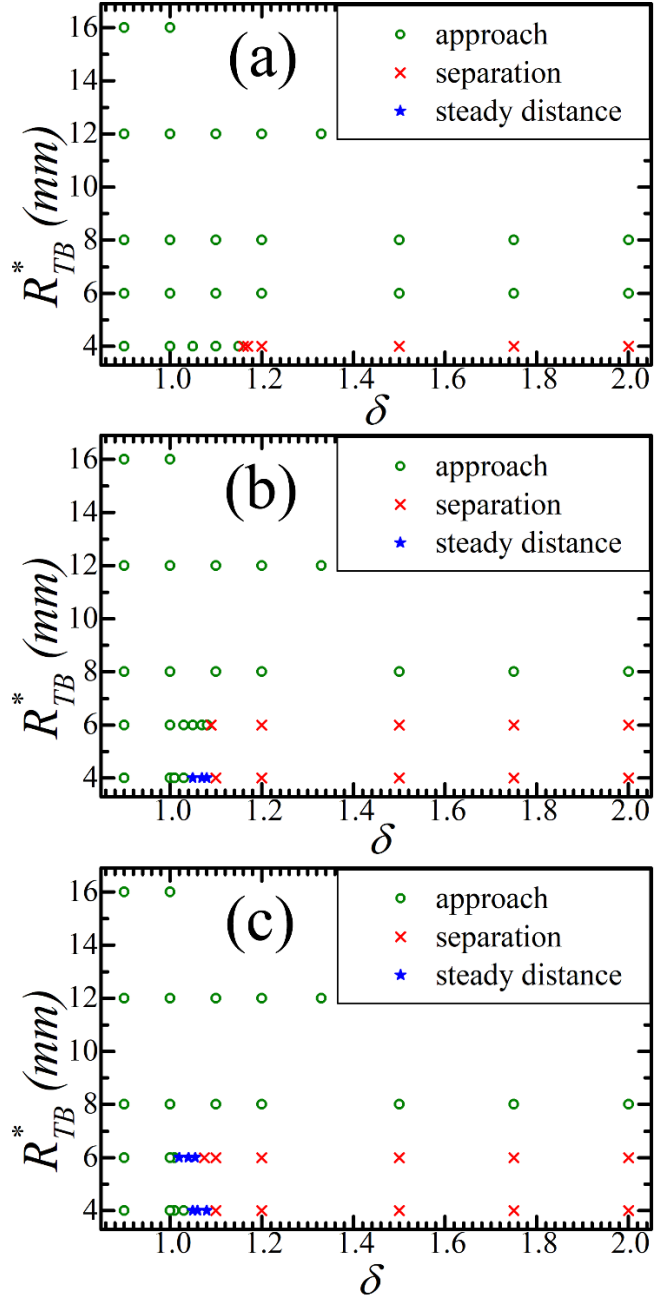

FIG 7 Map of regimes for combinations of $R_{TB}^*$ and $\delta$ at (a) $d_o = 4$, (b) $d_o = 5$ (base value), and (c) $d_o = 6$.

Next, we will focus our attention on pairs with $R_{TB}^* \leq 6\,mm$, starting from $R_{TB}^* = 4\,mm$. Close examination of bubble pairs with $R_{TB}^* = 4\,mm$ and any relative size in Fig. 7 (a) reveals that they do not attain a steady distance, in contrast to TBs of the same size in Fig. 7 (b) and (c) where they do. In Fig. 7 (a), we see the direct transition from *approach* to *separation* as $\delta$ increases. The short-range attractive dynamics is enhanced when the initial separation distance is smaller and leads to *approach* of the pair for $\delta$ up to just above unity. On the contrary, when $\delta$ is larger, the larger buoyancy of the LB separates the pair. In Fig 7 (b), where the value of $d_o$ is a little larger, the steady separation distance cases arise for pairs of $(\delta, R_{TB}^*)$ for which approach occurs in Fig. 7 (a). Hence, there must be an intermediate critical value of the initial separation distance for which the response changes from *approach* to *steady separation distance*. Our simulations reveal that this threshold is in the range $4 \leq d_o \leq 5$. Increasing further the initial separation distance in Fig. 7 (c), the transitions from *approach* to *steady distance* to *separation* as $\delta$ increases persist as in Fig. 7 (b). It should be anticipated that the same behavior along the line $R^*_{TB} = 4\ mm$ shown in Fig 7 (b) an (c), should also be observed for any other larger value of the initial separation distance, e.g., $d_o=10$.

Now we turn to pairs with $R^*_{TB} = 6\ mm$. In Fig. 7 (a) we observe only bubbles approaching, because the attractive dynamics are dominant in this $d_o$. We increase the initial separation distance by unity and in Fig. 7 (b) we observe a direct transition from *approach* to *separation* as $\delta$ increases. In this $d_o$, the short-range attractive dynamics are still relevant, and we see approach at small values of $\delta$, but separation for larger values. We increase $d_o$ further to 6 and we see also *steady separation distance* cases arising in Fig. 7 (c). This means that for $5 \leq d_o \leq 6$, there is a critical value above which the area *with steady separation distances* is formed. Comparing this critical value with the earlier one for $R^*_{TB} = 4\ mm$, we deduce that larger TB, having even larger leading bubbles in their front ($\delta > 1$), require larger initial separation distance to reach a *constant separation distance*. This is a result that stems from the size of the wake each LB generates, taking into consideration that the LB is larger than the TB. We will further examine how the characteristics of the wake affect the interaction in section IV.C.2. Most importantly, we will investigate rigorously in section IV.D how and why even the transient appearance of a "weak" negative wake acts as a retarding factor on the approach of the bubble pair. Regarding the steady separation distance, if we increase the LB by 0.1 mm the steady distance increases by 6 to 7 mm.

Pairs of large enough bubbles so that the ratio of buoyancy to viscous forces is significant, i.e. $R^*_{TB} \geq 8\ mm$, show only *approach*, regardless of the values of $d_o$ used in Fig. 7. However we have verified through simulations, the results of which are not provided in this work for the sake of conciseness, that these pairs indeed separate at larger $d_o$, $d_o = 8$. We are confident that the *steady separation distance* is not formed by these pairs even at larger initial distances because the stabilizing role of elasticity fades away as inertial effects are amplified. Instead, the balance shifts towards the interplay between inertial, viscous and capillary forces. Based on the work of Zhang et al. [47], capillarity contributes in a major way to the attractive dynamics of tandem bubbles, when $Bo$ is larger, e.g. in larger bubbles.

### C. The steady separation distance

For the examined material, our simulations predict that the establishment of a *constant separation distance*, $d_s$, takes place in a very limited parameter range. A better understanding of the required conditions for it may assist in proposing actions for prolonging the stability of emulsions, a very important technological problem. So, it is worthwhile investigating further this sensitive dynamic balance.

In this section we compare its value between cases in which it arises. We also make a useful comparison of various features of each bubble in the pair with those of an isolated (single) bubble (SB) of the same size. We will focus mainly on two cases that exhibit *steady separation distance*: case (G) with $R^*_{TB} = 4\ mm$, $\delta = 1.06$ at $d_o = 5$ and case (H) with $R^*_{TB} = 6\ mm$, $\delta = 1.04$ at $d_o = 6$. The corresponding parameter values are given in Table IV. Bubbles with $R^*_{TB} = 4\ mm$ were examined in section IV.A.1 also, so the dimensionless numbers coincide with those in Tables II and III but are repeated below to facilitate comparison.

Table IV. Parameter values for cases (G), (H), highlighting the size ratio

| Case | $R^*_{TB}$ | Ar | Bo | Eg | Bn | $\delta$ |
|---|---|---|---|---|---|---|
| (G) | 4 mm | 3.61 | 2.15 | 0.97 | 0.12 | **1.06** |
| (H) | 6 mm | 5.95 | 4.84 | 1.46 | 0.082 | **1.04** |

### 1. Effect of $d_o$ on $d_s$

We start our analysis with the pair of case (G) and then we vary the initial separation distance to evaluate how the steady distance is affected. In Fig. 8 we show the evolution of the separation distance for five initial separation distances. In Fig. 8 (a), for $d_o < 5$, namely $d_o = 3$ and 4, the bubbles approach each other. The near-field attraction that the bubbles experience is intense. The LB has limited space at its rear pole and cannot deform to a hydrodynamically favorable shape, hence the TB catches up with it easily, exactly like in the interaction mechanism described for equal bubbles in [7]. In Fig. 8 (b), when $d_o \geq 5$ the behavior changes radically. All the plots arrive at the same value of $d_s = 6.8$. This is in qualitative accordance with the study of Bot *et al.* [42], who reported that the same pairs of equal spheres settling coaxially in a Boger fluid from different $d_o$, end up having the same steady distance. In our case, as we mentioned in the previous section, we can identify the existence of a threshold $d_o$ value below which the bubbles approach, whereas above it the pairs arrive at a unique *steady distance* $d_s$ for each pair.

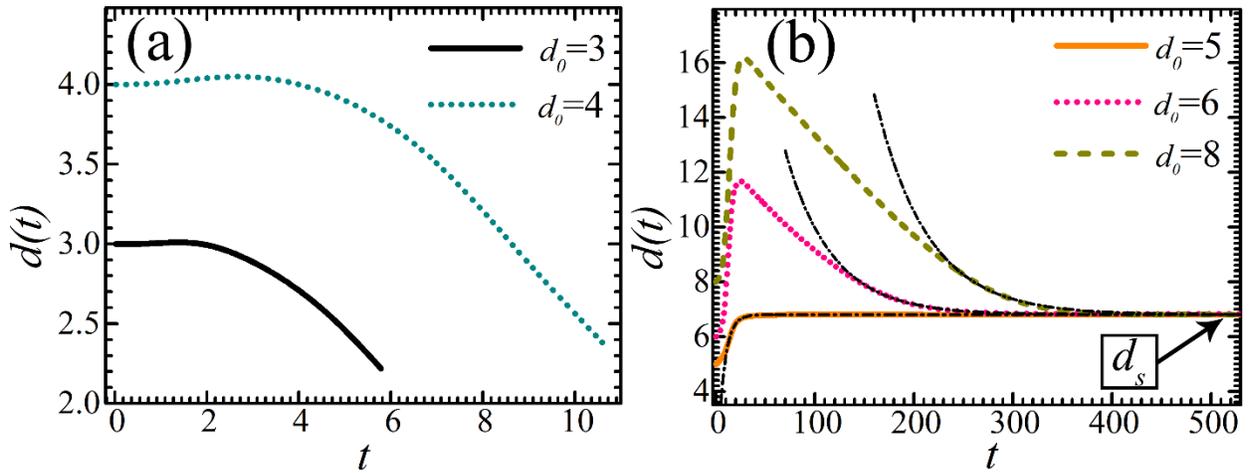

FIG 8 Time evolution of the separation distance for (a) $d_o = 3$, $d_o = 4$ and, (b) $d_o = 5$, $d_o = 6$ and $d_o = 8$ along with their corresponding fitted exponential decay functions with black dash-dot lines reaching $d_s$, given in Eq (19).

In Fig. 8 (b), we observe that for $d_o = 5$, (case (G)), $d_s$ is reached without the separation distance going through a global maximum first. We avoid using the term "monotonically" because as we saw in section IV.A.1, the separation distance of case (C) goes through a local minimum before eventually arriving at $d_s$. On the contrary, there is a global maximum for $d_o = 6$ and $d_o = 8$. These two local maxima arise when the TB has just formed the inverted teardrop-shape, a very hydrodynamically favorable shape, leading to the decrease of resistance exerted on it, and closing the gap to the LB. All three curves show an exponential decay to the steady value. The fitted expressions for the exponential decay are:

$$d_o = 5 \rightarrow f(t) = 6.8 - 7.2 \exp\left(-\frac{t}{6.5}\right)$$
$$d_o = 6 \rightarrow f(t) = 6.8 + 26.4 \exp\left(-\frac{t}{47.2}\right) \quad (19)$$
$$d_o = 8 \rightarrow f(t) = 6.8 + 166 \exp\left(-\frac{t}{52.8}\right)$$

The way the velocities evolve during the interaction of each pair shows a very interesting interplay. In Fig. 9 (a), we see that the common terminal velocity of all TBs is significantly larger, approximately 15%, than the velocity of the isolated bubble of the same size. This is expected due to material softening, which decreases the drag on the TB [7]. All curves of the TB velocities converge to the same value which is equal to the converged velocity of the LB shown in Fig. 9 (b). However, the transient period between the undershoot and the steady conditions differs. When $d_o = 5$, the velocity increases after the undershoot and damps asymptotically to the steady value. On the other hand, when $d_o = 6$ and $8$, the velocity increases, goes through an overshoot and then decays to the steady value. The velocity for $d_o = 6$ decays faster than that for $d_o = 8$. This is reflected also in the way the separation distance drops in Fig. 8. Up to the overshoot the $d_o = 6$ and $d_o = 8$ curves almost coincide. The same reduction of drag is exerted on them due to the presence of the LB but after that, a factor is slowing down both, and with a phase difference. The same factor does not allow the velocity for $d_o = 5$ to create an overshoot before reaching the same value with the other two curves. This factor is a negative wake generated at the back of the LB.

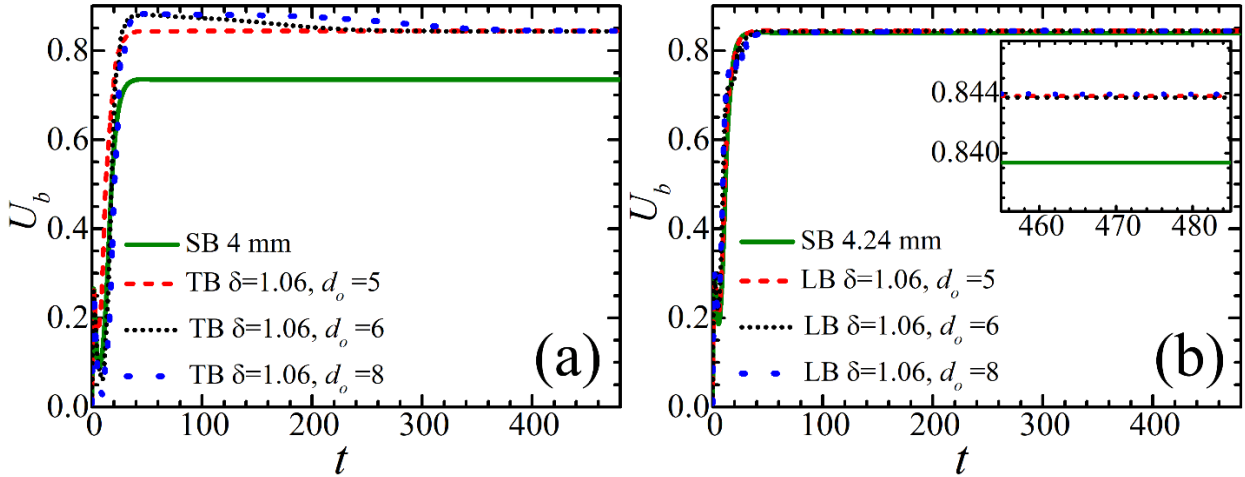

FIG 9 Time evolution of the bubble velocities of (a) the TB of each pair along with the velocity of the SB of the same size, (b) the LB of each pair along with the velocity of the SB of the same size.

In Fig. 9 (b) we see that the common terminal velocity of all LBs is slightly larger than the one of the single bubble of the same size. The difference is better shown in the inset of Fig. 9 (b). Our prediction on the velocity increase of the LB matches the observation by Bot *et al.* [42], who saw an increase in the velocity of the leading sphere of the pair settling in Boger fluid prior to the formation of *the steady separation distance*. Another very interesting investigation is the work by Gheissary *et al.* [60]. Their experimental results showed that spheres positioned initially in a side-by-side formation in Carbopol rearrange and eventually settle coaxially in a *steady separation distance*. This underlines that the *steady separation distance* we predict for bubbles has been observed experimentally in yield-stress fluids with elasticity. More importantly, it shows that the *steady separation distance* in elastic yield-stress fluids is a stable physical solution even when the initial conditions differ from the coaxial placement of the bodies. Finally, it makes relevant the assumption of axisymmetric arrangements in elastic yield-stress fluids. We should note that they studied interaction of spheres rather than bubbles in elastic non-Newtonian fluids and, to our knowledge, very few experimental works have been conducted on the interaction of bubbles in viscoelastic fluids [31][61]. Although the two problems (spheres and bubbles) differ significantly in terms of the physical state and the deformability of the moving bodies, as well as the applied boundary conditions on their surface, they are connected through the elastic behavior of the surrounding medium, which creates a Weissenberg-dependent elastic wake behind the moving bodies [24]. We will further analyze the influence of the wake on the *steady separation distance* in section IV.C.2. Apart from the experimental study of Bot *et al.*, *steady separation distance* of tandem spheres has been predicted and attributed to elasticity by Pan *et al.* [62], who modeled the medium as Oldroyd-B and FENE-CR fluids.

### *2. Factors contributing to the steady separation distance*

In all cases resulting in constant separation, the LB is only slightly larger than the TB, so, in principle, it should attain a larger rise velocity, which would split the pair. This velocity difference is moderated by material softening by the LB and the limited distance between the two bubbles, which often prevents the LB from assuming the inverted tear drop shape, whereas the TB does attain this hydrodynamically favorable shape. However, the decreased drag on the TB and shape adjustments do not seem to be sufficient, because they take place in several other cases, leading either to bubble *separation* or *attraction*, instead of the *constant separation distance*. Somehow the desired dynamic interaction must be stabilized further. We have identified two more reasons for this. The first one is the reduction of the acceleration or the deceleration of the TB, depending on $d_o$, as seen in Fig. 9 (a), and the second one is the small velocity increase of the LB. The former is achieved by the formation and preservation of a distorted negative wake behind the LB, appearing as a recirculation, and the latter by the modification of the stresses exerted on the LB, due to the presence of the TB. Together all these effects lead to the matching of the velocities between the LB and the TB, resulting in a constant distance between them.

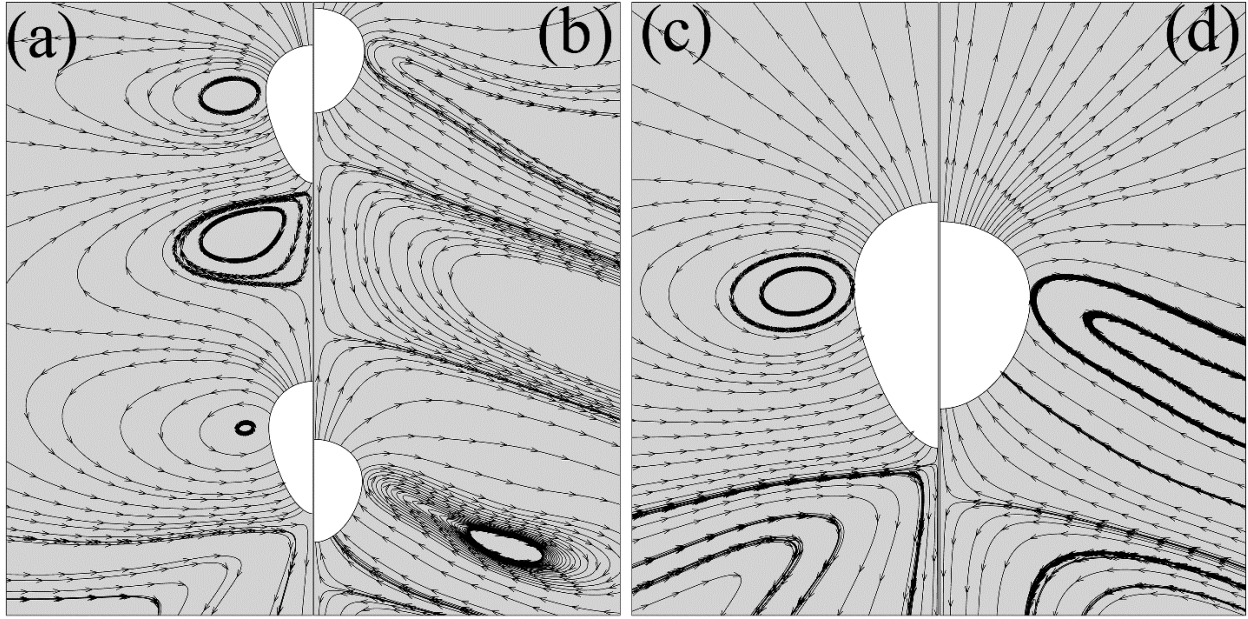

FIG 10 Streamlines at steady conditions of (a) the pair of case (G), (b) the pair of case (H), (c) a SB of the same size as the LB of (a) (4.24 mm), and (d) a SB of the same size as the LB of (b) (6.24 mm).

In Fig. 10 (a) we show the streamlines of the pair for case (G). We can clearly distinguish three stagnation points. The first one, just behind the LB, is the one that appears even for a single bubble that generates the negative wake behind it (Fig. 10 (c)). Similarly, the second one is just behind the TB. There is, however, an additional stagnation point, between the two bubbles that needs further investigation. The same triplet of stagnation points is observed in Fig. 10 (b) for the pair of the case (H).

The generation of a negative wake at the back of a buoyant body has been investigated thoroughly and has been attributed to the interplay between shear and extensional viscoelastic stresses and material thinning [24][25]. Soft-jammed systems, like Carbopol, satisfy these necessary conditions, hence the negative wake is present behind a buoyant body in an elastic yield-stress fluid. Indeed, in the case of a single settling sphere, it has been observed experimentally in Carbopol solution [27][28] and predicted via simulations for EVP materials [29]. Similarly, in the case of a single rising bubble, the negative wake has been observed experimentally in an EVP gel [22] and recently predicted numerically [15].

The negative wake at the back of a single rising bubble appears during its unobstructed flow. It is indicative of the velocity and stress field around the bubble. Positioning a second bubble at the wake of the original one interferes with this original flow field, which adjusts accordingly. We should mention the perception that the negative wake is one of the factors contributing to the *separation* between coaxially settling bodies in viscoelastic shear thinning fluids [39]. This factor, however, is not sufficient for this *separation*, as shown in [40][43][63]. The limited effect of the negative wake is confirmed also in the experimental study of Verneuil *et al.* [40] in which two stagnation points are shown between a pair of coaxially settling equal spheres in a shear thinning viscoelastic fluid. Nevertheless, eventually the spheres approach, the negative wake behind the leading sphere disappears and they touch each other.

Returning to Fig. 10 (a) and (b), both LBs have an inverted tear drop shape, which is more pointed in the smaller one, and just behind them the flow field exhibits a negative wake. Interestingly here, their shapes are quite similar to the isolated bubble of the same size, which also exhibits a negative wake, as shown in Fig. 10 (c) and (d), respectively. The negative wakes in the isolated bubbles have the usual shape of a cone around the axis of symmetry, which includes fluid moving in the opposite direction than the bubble. In the presence of a TB the original negative wake may modify the velocity of the TB, which should have increased because of material softening. Clearly these two phenomena affect the velocity of the TB in the opposite direction. Simultaneously, the rising TB displaces material ahead of it towards the LB. This material and the material flowing downwards in the negative wake of the LB collide and create a second stagnation point in the space between the two bubbles. This "collision" decreases the intensity of the negative wake behind the LB in comparison to the one behind the isolated bubble (compare Fig. 10a with Fig. 10c), reducing the upward thrust towards the LB and possibly its velocity for the TB to catch up. The existence of two stagnation points requires sufficient initial space between the bubbles. On the other hand, generation of bubble interaction requires that this distance is not too large. These two limitations establish the range of initial distances leading to a dynamically *constant separation distance*. We could say that the negative wake is slightly repulsive, but not enough to maintain a steady distance between bubbles of similar size. We will delve deeper into this topic in the next section. In all cases that exhibit *approach* in Fig. 5 and Fig. 7 either no negative wake is developed behind the LB, when $d_o$ is small or it is developed and then suppressed, when $d_o$ is larger.

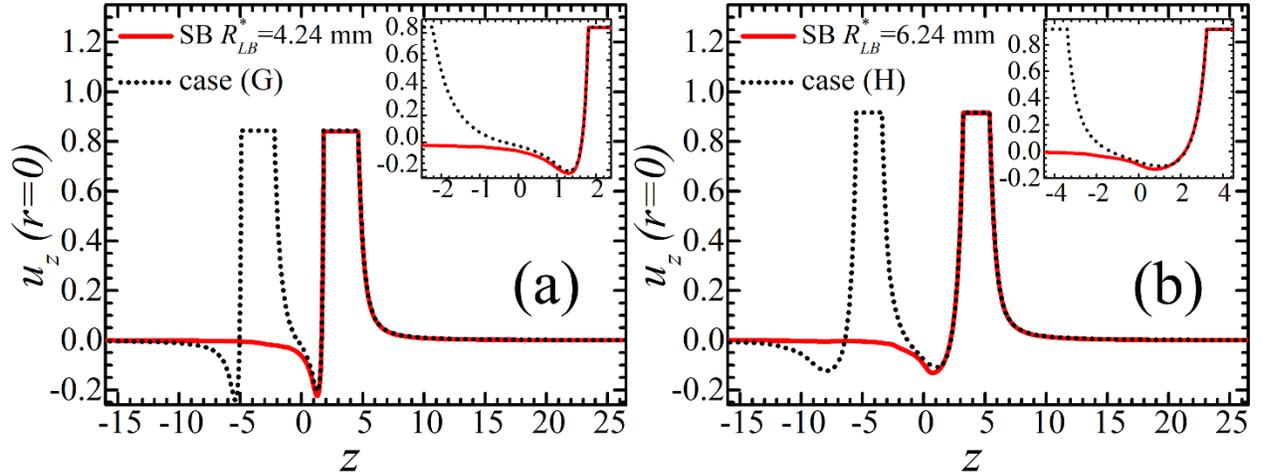

FIG 11 Axial velocity along the axis of symmetry after reaching a constant separation distance for (a) case (G) and for the SB of same size as the LB, and (b) case (H) and for the SB of same size as the LB. The inset of each figure shows an expanded view in the z-direction of the intermediate space between the bubbles.

In Fig 11 we show the axial velocity along the axis of symmetry for cases (G) and (H). Each plateau corresponds to the area occupied by a bubble, hence the steady velocity. In both panels the *steady separation distance* has been reached, so the velocity values of the two plateaus are the same. The profile of the velocity around the LB matches the one of the isolated bubble up to the position of the minimum velocity, both in Fig. 11 (a) and (b). Beyond the position of the minimum, the velocity for the single bubble asymptotes to zero in the far field, while in the case of the pair it rises to reach the positive value of the TB. In both panels we observe that the minimum velocity behind the LB is slightly larger than the one behind the corresponding isolated bubble, because of the thrust by the TB. On the other hand, the position of the minimum is the same for the LB and the SB, meaning that a specific distance behind the LB is required to form the negative wake. This explains why there is a critical distance below which the bubbles approach each other. A smaller distance would not allow the negative wake to form, i.e. the slight repulsion it may generate does not arise, and attraction prevails, as shown in Fig. 8 (a).

Moreover, at the inset of Fig. 11 (a), we observe that the distance between the minimum velocity position and the rear pole of the LB for case (G) is significantly smaller compared to the respective distance shown at the inset of Fig. 11 (b) for case (H) (0.52 vs 2.43 respectively). This derives from the axial velocity of the isolated bubbles shown in Fig. 11 and it is also evident in Fig. 10 (c) and (d) by inspecting the streamlines and the positions of the stagnation points. The shifting of the wake further downstream is a well-documented phenomenon for a settling sphere in shear thinning viscoelastic solutions. Arigo and McKinley [24] indicate that this is an effect of increased inertia. Inertia of the fluid is more prominent in the case of larger bubbles in case (H) compared to those in case (G) shifting the wake farther downstream. The same conclusion can be made by comparing the values of the Archimedes number in Table IV. This explains why a pair with $R_{TB}^* = 6\ mm$ shows a larger critical value of the initial separation distance, for *steady separation distance* to be expressed, as discussed in relation to Fig. 7. The wake needs more space to develop between the bubbles. We should underline that the presence of the negative wake between a pair of bubbles is a necessary, but not a sufficient condition. The combined buoyancy of LB-TB pair should be such that it balances the drop of drag on the TB [7].

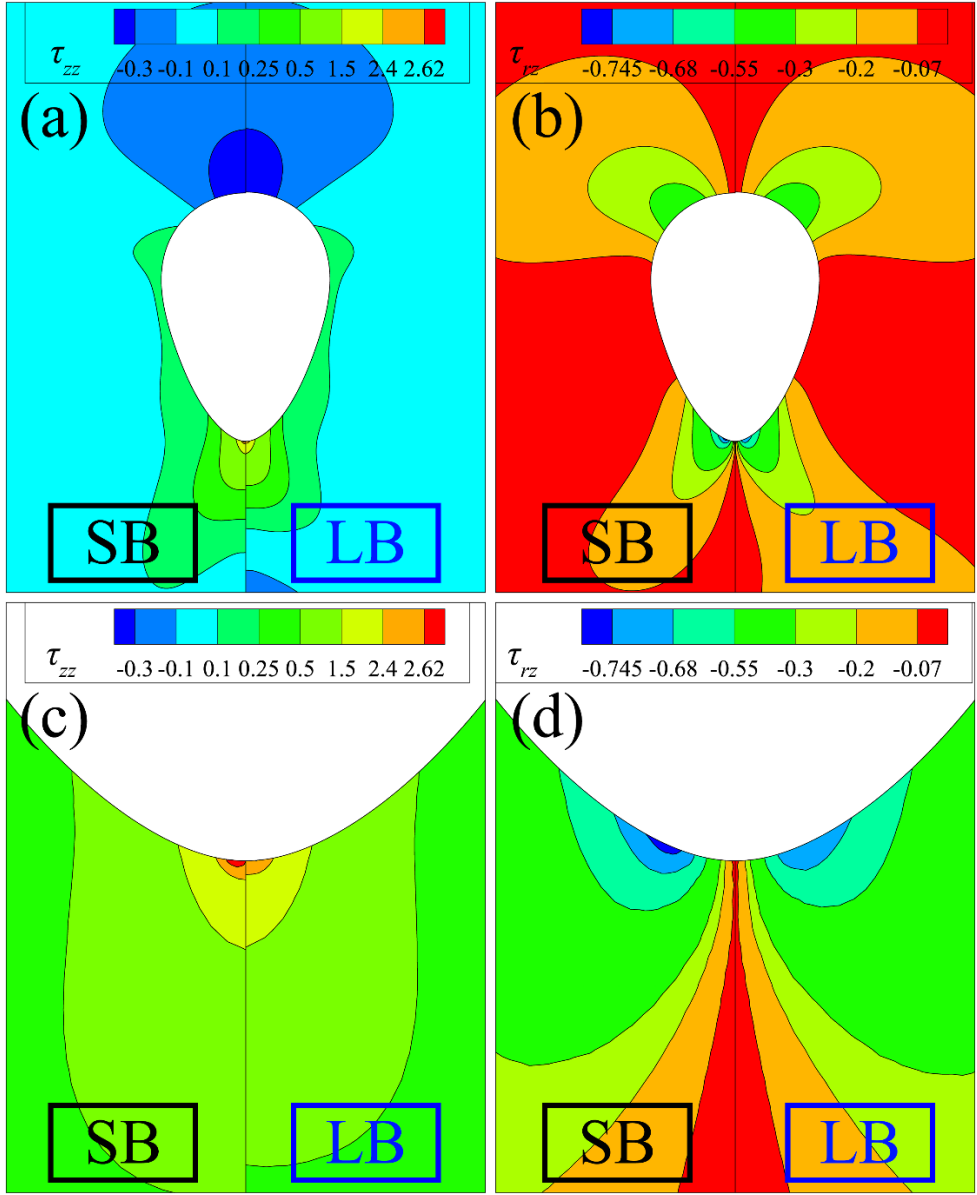

FIG 12 Comparison of stresses of an isolated single bubble (SB) of $R^*_{TB} = 4.24\ mm\ (= 1.06 * 4\ mm)$ to the left and the LB of case (G) to the right of each panel showing (a) the shear stresses, (b) the normal axial stresses, and a magnification at the rear pole of (c) the shear stresses, and (d) the normal axial stresses.

Next, we examine the mechanism leading to the velocity increase of the LB. The LB from case (G) will be used to demonstrate the small suppression of stress with respect to the SB of the same size. In Fig. 12 we show the stress field around a single isolated bubble denoted as SB and the LB of the pair when steady state has been achieved. The bubble shapes coincide. In Fig. 12 (a) the contours of the normal axial stresses are very similar in the front and side of each bubble, but at the back the contours around the LB are squeezed in comparison to the ones around the SB due to the presence of the TB. On the contrary in Fig. 12 (b) the contours of the shear stresses around the LB can be clearly seen to extend farther back to the TB as we commented in section IV.A.1. At first sight, no particular difference exists close to the surfaces of the LB and the SB. However, if we magnify the area of the rear pole, we discover in Fig. 12 (c) that the normal axial stress behind the SB is locally slightly larger than the one behind the LB. The same holds for the shear stress in Fig. 12 (d). Clearly the presence of the TB has modified the stresses at the back of the LB, reducing the drag on it and slightly increasing its velocity. Of course, we do not expect the velocity of the LB to be significantly influenced since the effect is only restricted to its rear pole. The increase of the velocity is approximately 0.006, as reported in Fig. 9 (b). Nevertheless, this increase happens simultaneously with the velocity change of the TB, matching the two perfectly. If this small increase did not occur the pair would slowly but steadily approach.

### D. Slowing effect of the leading negative wake on the TB

We conclude our analysis on the impact of the negative wake on the dynamics of a pair of bubbles by examining the time evolution of the kinematics. This analysis should assist alleviating the divergence on its effect on an interacting pair of bodies, which is seen in the current literature [39][40]. We use a pair of bubbles with $R_{TB}^* = 6$ mm and $\delta = 1.0$ (equal bubbles) in an initial separation distance $d_o = 6$ to demonstrate and explain the slowing effect of the negative wake on the TB and the inability of equal bubbles to form a steady separation distance. The corresponding dimensionless numbers are given in Table IV.

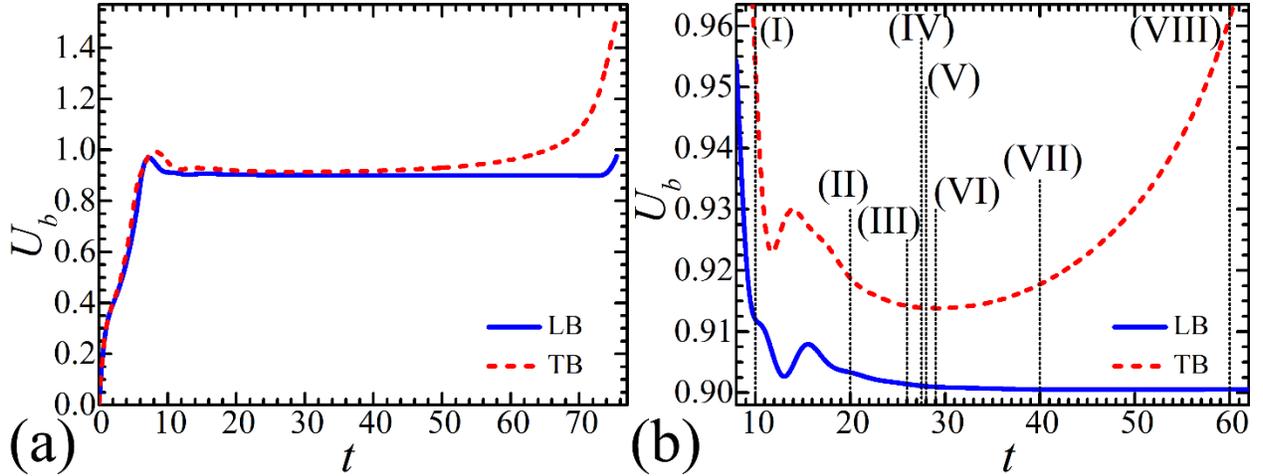

FIG 13 Time evolution of bubble velocities for the LB and TB (a) from the beginning of the motion up to the point the bubbles nearly touch, (b) expanded view for intermediate times. With Latin numerals we denote the time instants of the snapshots we show in Fig. 14. The time instants are (I) 10, (II) 20, (III) 26, (IV) 27.5, (V) 28, (VI) 29, (VII) 40, and (VIII) 60.

In Fig. 13 we show the time evolution of the velocities of both bubbles before they come in contact. Since the bubbles have the same buoyancy, it is expected that they will approach; see discussion in section IV.B and in [7]. In Fig. 13 (b) we show a magnification of Fig. 13 (a), including specific time instants at intermediate times. In the time interval of the expanded area, the velocity of the TB firstly exhibits a local minimum at $t \sim 12$ and then a more pronounced and broader second local minimum at $t \sim 29$. We will focus on the second local minimum. The TB decelerates down to velocity 0.914 at time instant (VI) and then accelerates. The factor that decelerates the TB is the presence of a negative wake behind the LB, as explained in the following paragraph.

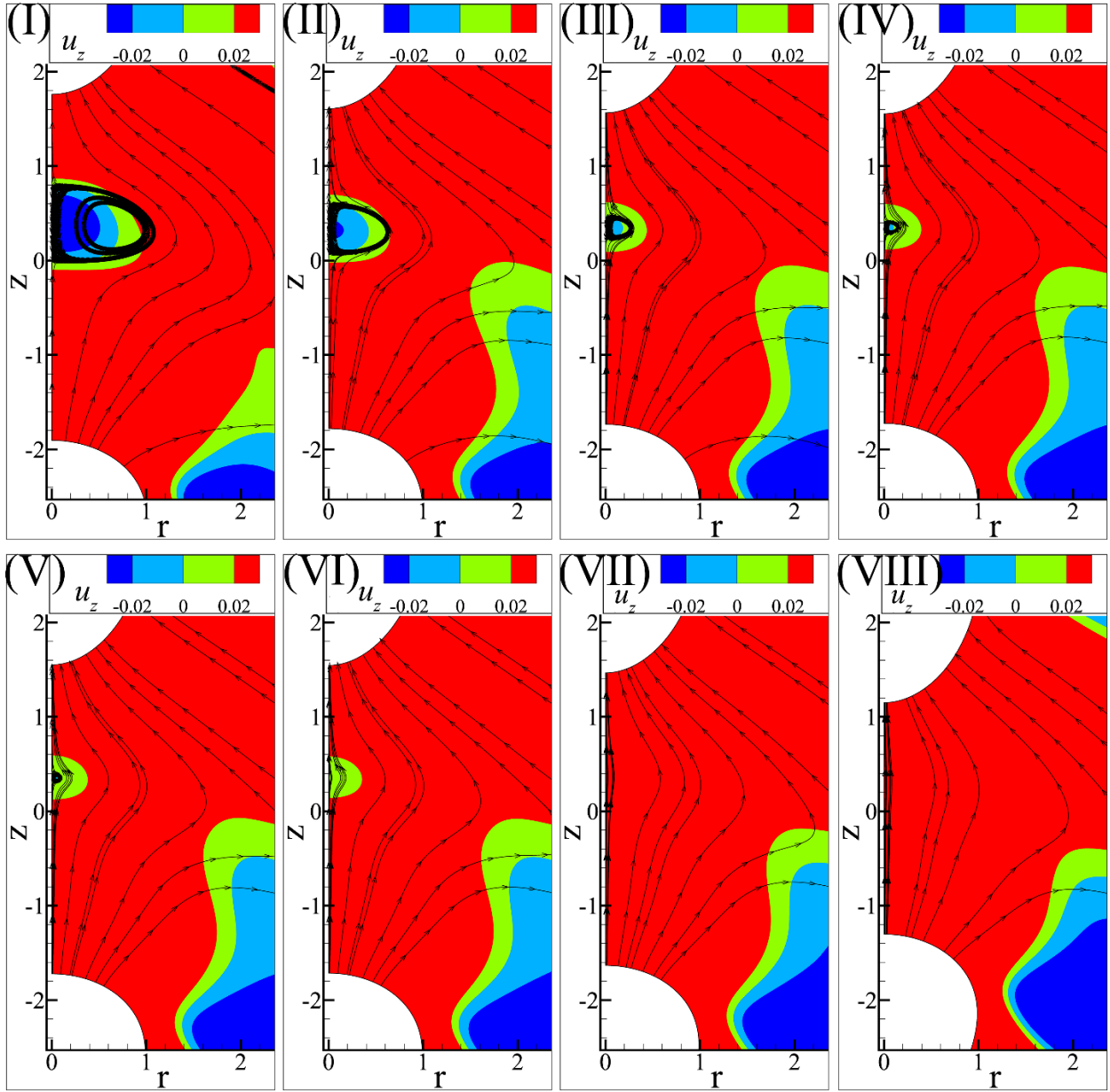

FIG 14 Axial velocity contours superimposed on the pathlines of the material in the intermediate region between the coaxial bubbles at time instants (I) 10, (II) 20, (III) 26, (IV) 27.5, (V) 28, (VI) 29, (VII) 40, and (VIII) 60.

In Fig. 14 (I) we see that a distorted negative wake exists behind the LB. It is delineated by two stagnation points at the axis of symmetry. We also show the axial velocity contours, which are negative (blue color) in the negative wake. Ahead and behind it the fluid moves upward. The dynamics around the TB has favored its motion compared to the motion of the LB up to time (I) because of all the mechanisms we have discussed that reduce its drag. So, the TB with its larger velocity approaches the LB. This sudden decrease of the space between the bubbles affects the negative wake and reduces its size and intensity at time (II). The rounded TB and the recirculation of the negative wake force the material undergoing displacement to follow curved streamlines. This increases the work needed to displace the material, and, hence, the drag exerted on the TB, decelerating it. Its velocity continuously decreases as time passes from (III) up to (V) with an increasingly smaller deceleration (the slope in Fig. 13 (b) is negative and gradually approaches zero). Nevertheless, the TB velocity remains larger than the LB velocity, maintaining their approach. Necessarily the negative wake shrinks further, as shown in Fig. 14 (III)-(V).

Finally, the negative wake disappears in Fig. 14 (VI), but some curved streamlines close to the axis of symmetry remain at its prior position. At time (VI) the TB reaches the local minimum of the velocity in Fig. 13 (b). Subsequently, in Fig. 14 (VII), the pathlines near the axis of symmetry align further with it allowing the TB to displace straight upward material from its front. Therefore, the TB increases its velocity again and the approach continues becoming even more pronounced in Fig. 14 (VIII), now that the pathlines are almost fully aligned with the axis of symmetry.

Based on this picture, the negative wake presents a source of resistance to the trailing body, whether it is a bubble, a sphere or a drop. The rheological behavior of the surrounding medium, then plays a crucial role on the degree of resistance the negative wake poses. In the present study, the response of the material is very soft, because it is both shear and extension thinning. That is one reason we cannot predict a steady separation distance for equal bubbles. The mechanisms reducing the drag on the trailing body prevail over the increasing resistance due to the negative wake. Instead, the steady separation distance is possible for slightly larger LB over TB, because the former attains a larger rising velocity avoiding the approach of the TB, although the drag on the TB drops, and generates a more intense negative wake that opposes more strenuously the motion of the TB. However, a coaxial and equally buoyant pair in a Boger fluid, that exhibits a bounded strain hardening behavior and no shear thinning, could achieve a steady separation distance.

### E. Parametric analysis on material properties

In this section we investigate the effect of the two main material properties of interest, namely the elastic modulus $G^*$ and the yield stress $\tau_y^*$. We delve into the dynamics of the system when the matrix is more elastic, $G^* = 24\ Pa$, than the base material, $G^* = 40.42\ Pa$, affecting only the value of $Eg$ at a given TB radius. Then, we examine the dynamics when the matrix is more plastic, $\tau_y^* = 7\ Pa$, than the base material, $\tau_y^* = 4.8\ Pa$, affecting only the value of $Bn$ at a given TB radius. We conduct simulations for various combinations of $(\delta, R_{TB}^*)$ to construct the maps for $d_o = 5$. The maps shown at previous sections denote that the more interesting results predicting steady separation distance appear for values of $\delta$ between unity and 1.10, therefore we will focus our simulations on this range of $\delta$.

#### 1. Effect of elasticity

We set the elastic modulus to 24 Pa keeping the rest of the base material properties the same, as shown in Table I, and we keep $d_o = 5$. We summarize our predictions in Fig. 15.

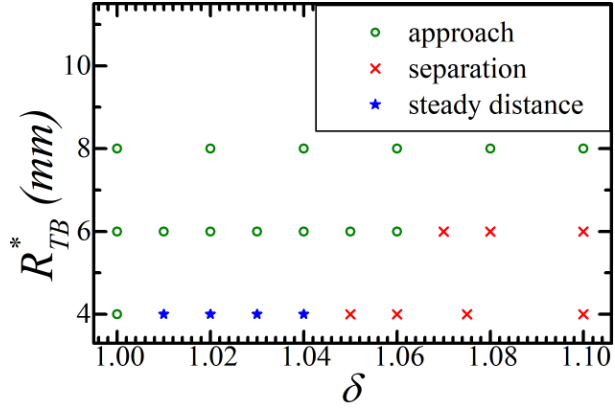

FIG. 15 Map of regimes in the $(R_{TB}^*, \delta)$ parameter plane for the more elastic matrix ($G^* = 24\ Pa$)

As we can see from Fig. 15, the increase in the elastic response of the material plays a moderately stabilizing role via the increased presence of the steady distance. This occurs for $R_{TB}^* = 4\ mm$, where the range of this outcome has increased compared to the base material shown in Fig. 7. Furthermore, the steady distance cases have been shifted towards lower values of $\delta$. Namely, in the base material they occur for $1.05 \leq \delta \leq 1.07$, whereas in the more elastic material for $1.01 \leq \delta \leq 1.04$. This shift to lower values of the size ratio is reasonable because the intensity of the mechanism inhibiting the approach, i.e. the negative wake, is amplified by elasticity. However, the negative wake cannot induce separation independently but requires the buoyancy of the LB being large enough to aid the tendency of LB to drift apart, hence the range is broadened. For $R_{TB}^* = 6\ mm$ the results are largely similar to those in the base material. Increasing $\delta$, the dynamics of the bubble pairs changes from *approach* to *separation*, with no *steady distance* being observed. No negative wake is predicted behind the LB for this size of TB due to the small space between the bubble pairs, which is in accordance with our analysis in section IV.C.2. For the larger $R_{TB}^* = 8\ mm$ only bubble approach is predicted.

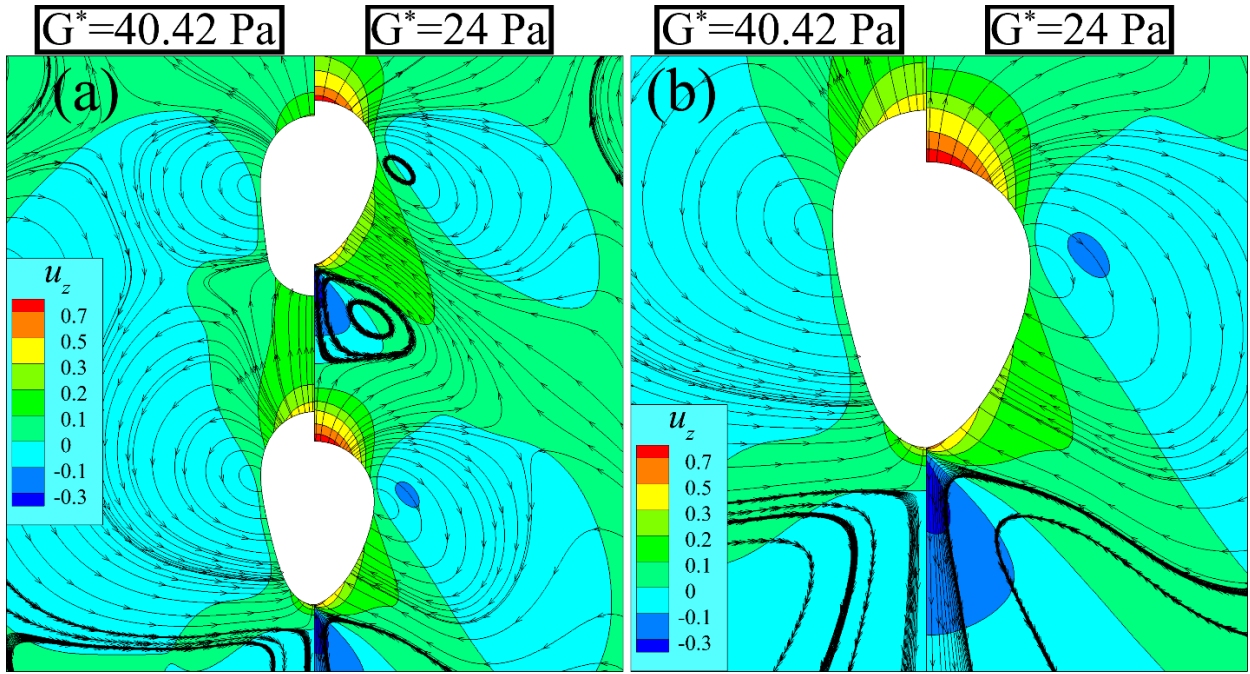

FIG. 16 The axial velocity superimposed to pathlines of the same bubble pair with $(R_{TB}^*, \delta) = (4\ mm, 1.03)$ at the same initial separation distance $d_o = 5$ in the base matrix to the left and in the more elastic matrix to the right of the snapshots, both at $t = 10$. In (a) the image shows both the LB and TB and in (b) the zoomed frame on the TB.

In Fig. 16 we see at $t = 10$ the snapshot of the same bubble pair rising at the base matrix to the left and at the elastic matrix to the right. The increased elasticity of the material leads to the generation and preservation of a negative wake in the intermediate space between the bubbles that keeps the pair to a *steady separation distance*, although the same pair approaches in the base matrix. Furthermore, we observe substantially larger velocities in the case of the elastic material due to the more hydrodynamically favorable shapes that develop faster. The intensity of the negative wake visibly increases and swifts the steady distance cases to smaller $\delta$.

### 2. Effect of yield stress

We set the yield stress to 7 Pa keeping the rest of the base material properties the same, as shown in Table I, and we keep $d_o = 5$. We summarize our predictions in Fig. 17.

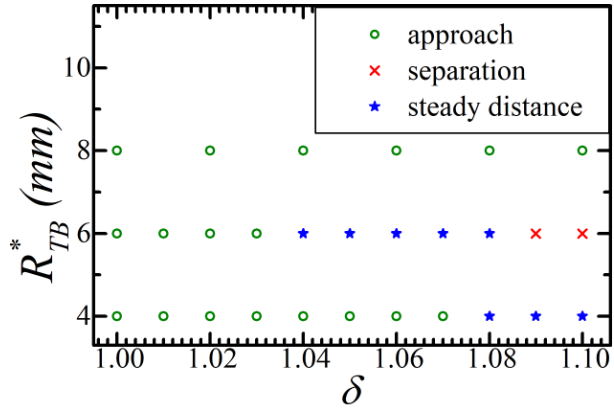

FIG. 17 Map of the regimes in the $(R_{TB}^*, \delta)$ parameter plane for the more plastic matrix $(\tau_y^* = 7\ Pa)$

In Fig. 17, we observe that the increase of the plasticity of the medium leads generally to significant increase of pairs with *steady separation distance*. For $R_{TB}^* = 4\ mm$, the range of the steady distance cases has remained the same compared to the base matrix, but it has shifted to larger values of $\delta$, $1.08 \leq \delta \leq 1.1$. Since the yield stress has increased and $Bn \sim 0.18$, which is above the entrapment condition for a single bubble of this size [15], the material becomes more brittle and the dynamics becomes slower. The elastic effects are less pronounced for values of $\delta \leq 1.07$. However, the residual stresses at the path of the leading bubble provide a route of reduced resistance and approach is encountered. On the other hand, for $\delta \geq 1.08$, the buoyancy of the LB is adequate to generate more intense dynamics that activate elasticity, which leads to the *steady separation distance*. The stabilization effect of plasticity is made clearer for $R_{TB}^* = 6\ mm$. With the base matrix, no *steady separation distance* is predicted for this size of TB, whereas in Fig. 17 the range of *steady separation distance* cases is quite wide, $1.04 \leq \delta \leq 1.08$. This occurs because the increase of $Bn$ brings the negative wake closer enough to the bubble, so that it fits in the space between the bubbles. To support this assertion, we demonstrate the pathlines of the same pair of bubbles in the base matrix and in the matrix with increased $\tau_y^* = 7\ Pa$ in Fig. 18.

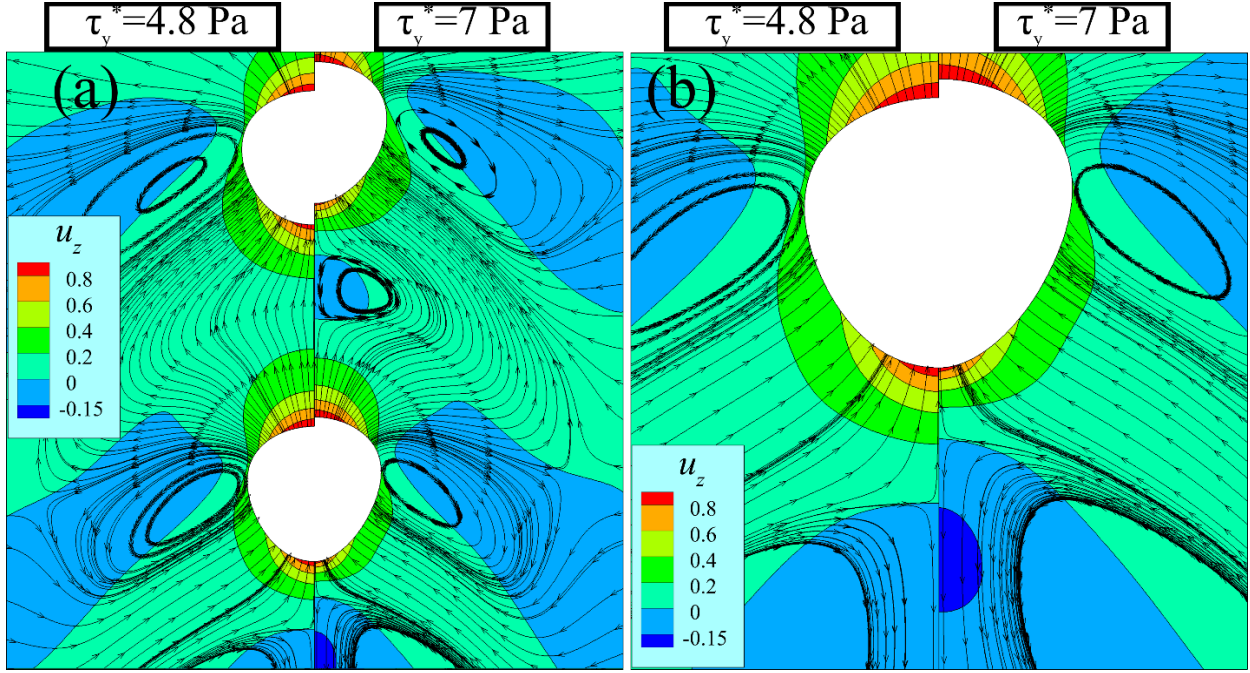

FIG. 18 The axial velocity superimposed to pathlines of the same bubble pair with $(R_{TB}^*, \delta) = (6\ mm, 1.06)$ at the same initial separation distance $d_o = 5$ in the base matrix to the left and in the more plastic matrix to the right of the snapshots, both at $t = 10$. In (a) the image shows both the LB and TB and in (b) the zoomed frame on the TB.

As we can observe in Fig. 18 (a), the same bubble pair, which is initially at the same separation distance, exhibits a distorted negative wake behind the LB in the more plastic matrix, contrary to the rising in the base matrix. The increase of $Bn$ shifts the stagnation point closer to the tail of the bubble, as it is clearly shown in Fig. 18 (b). The decrease of the distance of the stagnation point to the tail of the bubble allows the negative wake to fit in the intermediate space behind the LB and remain there, maintaining the steady distance. So, instead of bubble approach in the base matrix, we predict a steady distance in the more rigid material. For the larger $R_{TB}^* = 8\ mm$ only bubble approach is predicted, as in the base state.

## V. SUMMARY AND CONCLSUSIONS

We studied the interaction of coaxial bubbles of any relative size, while they rise due to buoyancy in an elastoviscoplastic material. We constructed maps summarizing the response of each pair at various initial separation distances and we recorded conditions leading to the three regimes: *separation*, *approach*, and *steady separation distance*. Bubble pairs with leading bubbles smaller than or equal to the trailing bubbles always approach due to the lack of excess driving force, given the drop in the drag of the trailing bubble. When the initial separation distance is small, the short-range dynamics is enhanced, leading to a tendency for approach. The same tendency for approach is predicted for bubbles large enough that the ratio of buoyancy over viscous forces is significant for all initial separation distances we tested. The sheltering effect generates favorable conditions for the leading bubble to close the distance.

On the contrary, when this ratio for the trailing bubble is moderate or less, all regimes can appear depending on the relative size of the leading bubble. A slightly larger leading bubble can lead the pair to form a *steady separation distance*. This stable configuration is attributed to elasticity, and it is caused by two additional features, the distorted negative wake in front of the trailing bubble and the moderation of stresses at the rear pole of the leading bubble. The negative wake acts as an obstacle to the trailing bubble and balances the reduced drag exerted on it due to the softening of the material. The confinement of stresses exerted on the rear pole of the leading bubble increases its velocity by a small amount. All these features together result in the two bubbles acquiring precisely the same velocity and, hence, preserving a *constant separation distance*.

The appearance of a *steady separation distance* for a pair of bubble sizes depends also on the initial separation distance. If the initial separation distance is below a threshold value which varies with the radius of the trailing bubble, *approach* of the pair occurs. If this critical value is surpassed, then steady distance may be attained. For any initial separation distance above the threshold, the bubbles will acquire the same steady distance, which is unique for each pair of bubbles. Bubbles of larger radii form a larger steady separation distance, which is attributed to the appearance of the stagnation point at a larger distance from the rear pole of the LB. This observation derives from the dynamics of a single bubble. The larger the ratio of buoyancy over viscous forces in a larger bubble shifts the negative wake further downstream and extends the wake. In other words, in order for the negative wake to fit between the bubbles of larger size, the minimum space needed is larger according to the unobstructed flow field of a single bubble.

We also discussed the obstructing role of the negative wake in front of the trailing bubble. To our knowledge, the present study is the first to report and explain the transient effect that a negative wake has on the interaction of moving bodies. This interaction depends on whether the material displaced from the front pole of the trailing bubble has to follow a straight path or a path of curved pathlines due to the presence of the negative wake in its front. This remark opens a discussion on the connection of the negative wake behind the leading body, which is inextricably linked to the rheological behavior of the material, with the outcome of the interaction of buoyancy driven tandem bodies. For instance, a material that allows the generation of a negative wake but shows a mild extension hardening behavior could possibly maintain the *steady separation distance* even if the two bodies have the same buoyancy.

Furthermore, we performed a parametric analysis on the effect of the two main material properties of interest, namely the elastic modulus and the yield stress. We detected that an increased elasticity of the surrounding matrix increases the presence of the *steady separation distance*, by broadening the range of bubble size ratios showing steady distance. This is in agreement with our conclusion on the necessity of elasticity for observing a steady separation distance. We also predict that an increased yield stress shows a similar effect on the stabilization of the steady separation distance. By increasing the latter property, larger bubble pairs exhibit a *steady separation distance* in the more plastic matrix, whilst *approach* or *separation* was exhibited by the same pairs in a less plastic matrix. This occurs because a larger buoyancy is needed to shatter the microstructure of the material, which now is stiffer than its base counterpart. The increased plasticity draws the negative wake behind the leading bubble closer to the rear of the bubble, enabling its size to fit in the confined space.

With this analysis we conclude the investigation on the interaction of tandem rising bubbles of any relative size in elasto-visco-plastic materials. In the future we will extend our analysis to the interaction of tandem drops in elasto-visco-plastic materials. Also, we plan to conduct 3D simulations of tandem bubble pairs in order to explore whether non axisymmetric solutions, such as those observed experimentally for bubble pairs rising in viscoelastic fluids [31], can appear in elasto-visco-plastic materials.

ACKNOWLEDGMENTS

This work was supported by the Hellenic Foundation of Research and Innovation, under Grant No. HFRI FM17-2309 MOFLOWMAT and by the European's Union's Horizon 2020 research and innovation programme under the Marie Skłodowska-Curie Grant Agreement N° 955605.